\documentclass{article}

\usepackage[a4paper, total={7in, 10in}]{geometry}
\usepackage[english]{babel}
\usepackage{pdfpages}
\usepackage{amssymb}
\usepackage{amsmath}
\usepackage{subcaption}
\usepackage{graphicx}
\usepackage{hyperref}
\usepackage{float}
\usepackage{authblk}
\hypersetup{
    colorlinks=true,
    linkcolor=black,
    filecolor=black,      
    urlcolor=black,
    citecolor=black
}
    
\usepackage{setspace}

\newcommand{\grad}{^{\circ}}
\newcommand{\cm}{\textrm{cm}}

\begin{document}


# Low-loss twist-tunable in-plane anisotropic polaritonic crystals

Nathaniel Capote-Robayna[1,2], Ana I. F. Tresguerres-Mata[3,6], Aitana Tarazaga Martín-Luengo[3,6], Enrique Terán-García[3,6], Luis Martin-Moreno[4,5], Pablo Alonso-González[3,6]*, and Alexey Y. Nikitin[2,7]*

[1]*Applied Physics department, Engineering school of Gipuzkoa, University of the Basque Country (UPV/EHU), Donostia-San Sebastián, 20018, Spain.*

[2]*Donostia International Physics Center (DIPC), Donostia-San Sebastián, 20018, Spain.*

[3]*Department of Physics, University of Oviedo, Oviedo, 30006, Spain.*

[4]*Instituto de Nanociencia y Materiales de Aragón (INMA), CSIC-Universidad de Zaragoza, Zaragoza, 50009, Spain.*

[5]*Departamento de Física de la Materia Condensada, Universidad de Zaragoza, Zaragoza, 50009, Spain.*

[6]*Center of Research on Nanomaterials and Nanotechnology, CINN (CSIC-Universidad de Oviedo), El entrego, 33940, Spain.*

[7]*IKERBASQUE, Basque Foundation for Science, Bilbao, 48013, Spain.*

*Corresponding author. Email: pabloalonso@uniovi.es, alexey@dipc.org*

## Abstract

Van der Waals (vdW) materials supporting phonon polaritons (PhPs) – light coupled to lattice vibrations –have gathered significant interest because of their intrinsic anisotropy and low losses. In particular, α-MoO$_3$ supports PhPs with in-plane anisotropic propagation, which has been exploited to tune the optical response of twisted bilayers and trilayers. Additionally, various studies have explored the realization of polaritonic crystals (PCs) – lattices with periods comparable to the polariton wavelength –. PCs consisting of hole arrays etched in α-MoO3 slabs exhibit Bragg resonances dependent on the angle between the crystallographic axes and the lattice vectors. However, such PC concept, with a fixed orientation and size of its geometrical parameters, constrains practical applications and introduces additional scattering losses due to invasive fabrication processes. Here we demonstrate a novel PC concept that overcomes these limitations, enabling low-loss optical tuning. It comprises a rotatable pristine α-MoO$_3$ layer located on a periodic hole array fabricated in a metallic layer. Our design prevents degradation of the α-MoO$_3$ optical properties caused by fabrication, preserving its intrinsic low-loss and in-plane anisotropic propagation of PhPs. The resulting PC exhibits rotation of the Bloch modes, which is experimentally visualized by scanning near-field microscopy. In addition, we experimentally determine the polaritons momentum and reconstruct their band structure. These results pave the way for mechanically tunable nanooptical components based on polaritons for potential lasing, sensing, or energy harvesting applications.

**Keywords:** hyperbolic polaritons, phonon polaritons, polaritonic crystals, twisted heterostructures, van der Waals materials, nanophotonics

## 1. Introduction

The discovery of polaritons supported by van der Waals (vdW) materials[1], [2] has sparked significant interest due to their potential for manipulating light on the nanoscale[3]. In particular, the propagation of in-plane anisotropic phonon polaritons (PhPs) is supported in certain crystal layers, such as calcite[4], α-MoO$_3$[5], V$_2$O$_5$[6], or bGO[7] among others. These materials exhibit PhPs with hyperbolic dispersion across specific frequency ranges, leading to exotic optical phenomena such as light canalization in twisted crystal bilayers[8], [9], [10], [11] and trilayers[12], [13], or twist-tunable nanoresonators[14]. These findings open possibilities for an active tuning of the polariton propagation through the twist angle between layers, which is the key feature of the emerging field known as twistoptics. Another interesting option for controlling polaritons on the nanoscale consists of constructing polaritonic crystals (PCs) – lattices composed of elements periodically spaced at distances comparable to the wavelength of polaritons – made in vdW materials, such as e.g. h-BN[15], α-MoO$_3$[16], [17], [18], monolayer graphene[19], [20] or twisted bilayer graphene[21]. PCs support ultra-confined Bloch modes[22], [23], [24] that can favor a topological funneling of PhPs[25]. Recently, PCs composed of hole arrays (HAs) in α-MoO$_3$ slabs have been suggested and studied, both theoretically[16] and experimentally[17], [18]. In these PCs, emerging Bragg resonances depend on the orientation of the lattice vectors with respect to the crystallographic axes of the vdW crystal layer. However, these HAs were etched directly in the anisotropic vdW layer, so that their geometry is not actively reconfigurable. Moreover, the fabrication of holes within the α-MoO$_3$ increases the losses due to severe scattering processes. Consequently, although the rotational dependence of the Bragg resonances has been showcased, achieving post-fabrication tunability remains a challenge.

In this work, we introduce a novel PC concept based on the fabrication of twistable α-MoO$_3$/metal heterostructures that overcomes the above limitations, enabling low-loss optical tuning. As depicted in Figure 1, it comprises a twisted pristine layer of α-MoO$_3$ on top of a periodic HA made in a gold layer. The lattice period matches the wavelengths of the PhPs in α-MoO$_3$ crystal on top of gold. This configuration allows for the active tunability of the angle between the lattice vectors and crystallographic directions by rotating the α-MoO$_3$ layer. Employing a theoretical approach, that incorporates full-wave simulations and an analytical approximation, we describe the formation of PhP band structure and excitation of Bragg resonances. Furthermore, we conduct near-field measurements for different rotation angles, managing to disentangle the contribution of individual PhP Bloch modes emerging in this configuration. With our combined theoretical approach, we also reconstruct the PhP bands from the near-field data, by extracting the wavelength of the PhPs along different directions in plane.

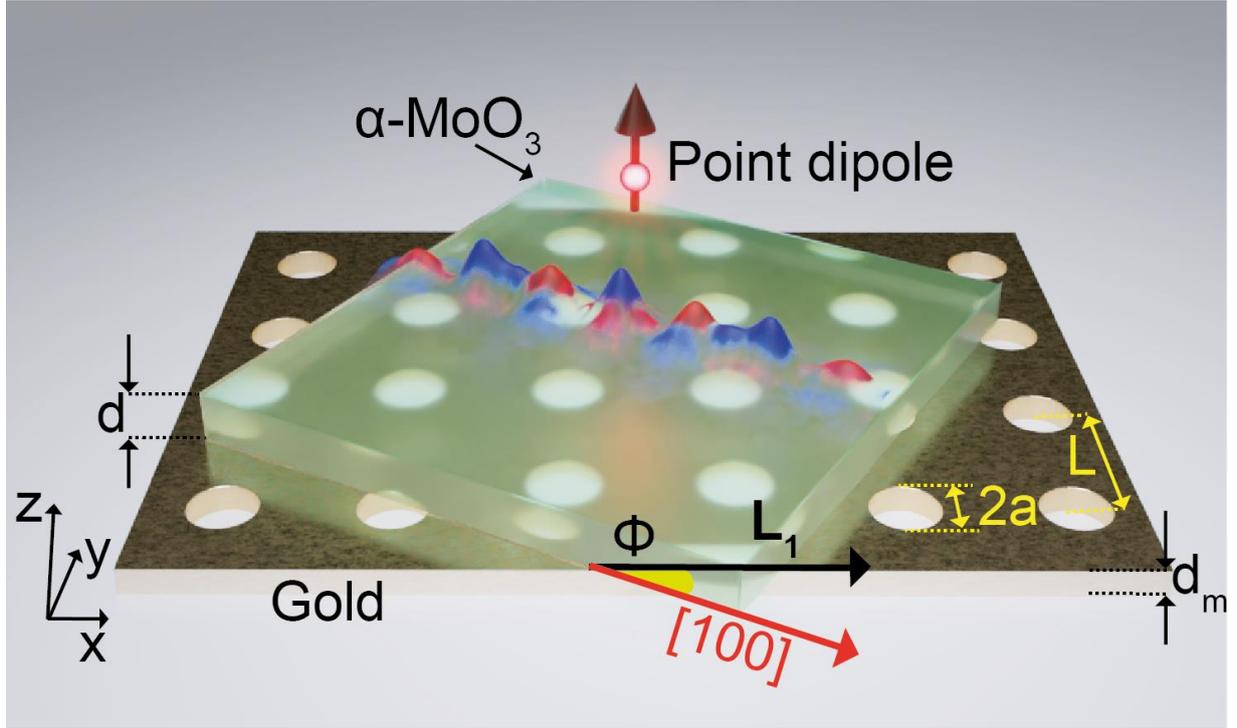

**Figure 1. Schematic of a low-loss twist-tunable polaritonic crystal: an in-plane anisotropic α-MoO₃ layer twisted on top of a HA made on gold.** The periodic lattice comprises a HA drilled in a gold layer with thickness $d_m$, hole radius $a$ and periodicity $L$. The anisotropic crystal layer, α-MoO₃, with thickness $d$, has its crystallographic axes twisted by an angle $\Phi$ with respect to the lattice vectors. To visualize the polaritons excited in the PC we modeled the distribution of the electric field (z-component), generated by a vertically oriented point dipole placed above the α-MoO₃ layer.

2. **Results**

In Figure 1 we show the schematics of our heterostructure representing a twisted PC. A pristine layer of α-MoO₃ (of thickness $d$) lies on a metallic film (of thickness $d_m$) with a periodic array of holes (of radius $a$, and filled by a material with permittivity $\varepsilon_h$) etched in it. Although our concept is valid for any periodic lattice with arbitrary lattice vectors $L_1$ and $L_2$, here for simplicity we focus on a square lattice ($L_1$ and $L_2$ are orthogonal and $|L_1| = |L_2| = L$). The α-MoO₃ layer is twisted by an angle $\Phi$ with respect to the $L_1$ lattice vector of the HA. Notice that the thickness of the gold film is assumed to be larger than the skin depth, so that the PhPs in the vdW layer are not sensitive to $d_m$.

Crystal layers of α-MoO₃ support PhPs within three different Reststrahlen bands[26] (RBs, range of frequencies defined between the longitudinal and transversal optical phonons frequency, LO and TO, respectively): $544.6 cm^{-1}$ – $850.1 cm^{-1}$ for RB1, $821.4 cm^{-1}$ – $963 cm^{-1}$ for RB2, and $956.7 cm^{-1}$ – $1006.9 cm^{-1}$ for RB3. In these RBs the isofrequency curves (IFC) – describing available in-plane momenta at a fixed frequency – can take different shapes. Thus, in RB1 and RB2 IFCs have hyperbola-like shapes, with vertices lying on the [001] and [100] crystallographic

directions, respectively. In contrast, PhPs with elliptical IFCs are supported in RB3. These highly anisotropic PhPs evolve into a series of modes, commonly designated as $Mn$, where $n \in \mathbb{N}$ indicates the quantization of the mode in the transversal (perpendicular to the slab) direction, displaying a lower propagation length and wavelength when increasing $n$. As in the mid-IR frequency range gold behaves as a perfect electric conductor (a "mirror"), only symmetrical modes could be excited in an α-MoO$_3$ layer on top of a gold film, so that the mode with the longest polaritonic wavelength is M1[14].

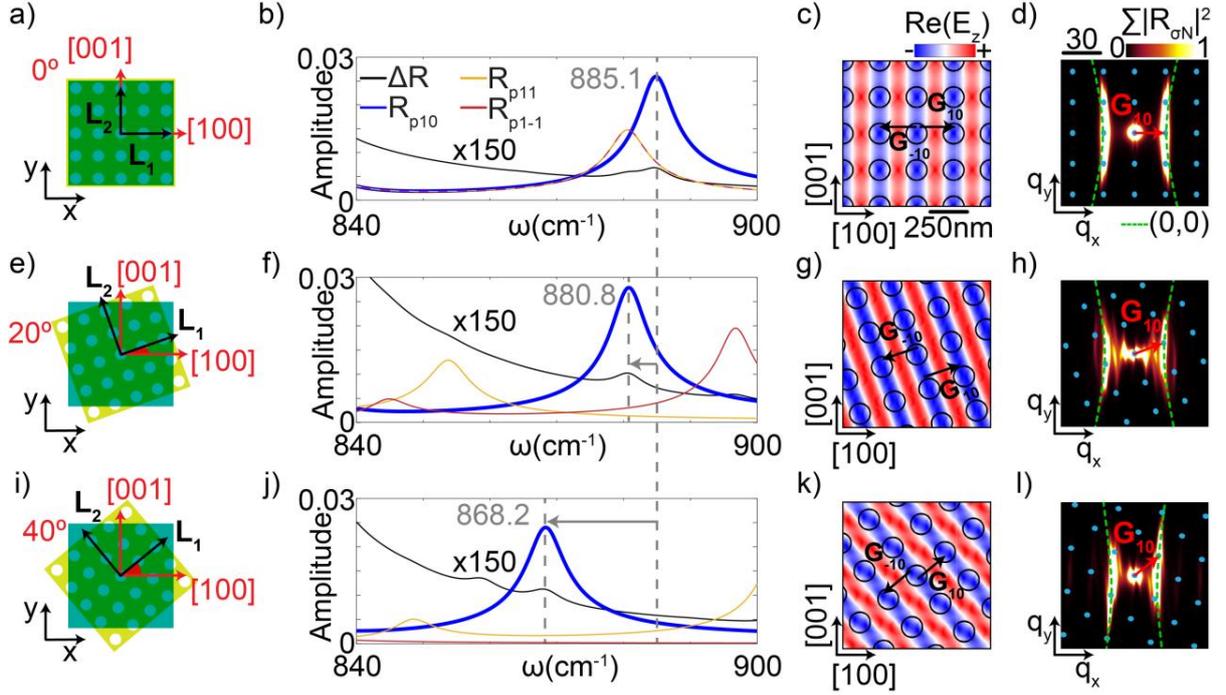

**Figure 2. Twist-tunable Bragg resonances in a PC. (a, e, i)** Schematics illustrating the top view of a twistable PC. The α-MoO$_3$ layer is aligned with the xy axis, whereas the gold HA (with the lattice vector basis, $L_1$ and $L_2$) undergo an anticlockwise twist of 0°, 20° and 40°, respectively. **(b, f, j)** Spectra of different Fourier field harmonics amplitudes for normal incident light linearly polarized along the [100] axis. Black line represents $\Delta R = R_b - R_{00}$, where $R_b$ is the reflection of α-MoO$_3$ on top of bare gold layer, and $R_{00}$ is amplitude of the (0,0) field harmonic. The blue, yellow and red lines represent $R_{p10}$, $R_{p11}$ and $R_{p1-1}$ (p-polarization components of the Fourier field harmonics). The shift of the resonance is indicated by a gray dashed line. **(c, g, k)** Electric field distributions for $\Phi = 0°$, 20° and 40°, at the $(\pm 1, 0)$ resonance frequencies, $885.1 cm^{-1}$, $880.8 cm^{-1}$ and $868.2 cm^{-1}$, respectively. The Bragg vector $\mathbf{G_{10}}$, perpendicular to the wave fronts and aligned to $L_1$ lattice vector, is indicated by the black arrow. **(d, h, l)** Colorplot (representing $\sum |R_{\sigma N}|^2$ for $\sigma \in \{s, p\}$) illustrates the IFC of the PC at the same twist angles and frequencies as in **(c, g, k)**. The reciprocal lattice and the Bragg vector, $\mathbf{G_{10}}$, are indicated by the blue points and the red arrow, respectively. IFC of the empty lattice for the diffraction order $(0, 0)$ is represented by the green dashed line.

To study and illustrate the emergence of tunable Bragg resonances in our twisted heterostructure, we calculate the field amplitudes of different diffraction orders (see Supplementary Material, Section 1 for details). The parameters of the lattice (period $L = 250nm$, and the hole radius $a = 55nm$, where the holes are filled by air, thus $\varepsilon_h = 1$) have been chosen to match the available wavelengths of the PhP mode in an α-MoO$_3$ slab of thickness $d = 70nm$ (in this work we will focus on the RB2). For simplicity, but without loss of generality, we assume that both the substrate and superstrate are air, while the thickness of the metal film is set to $d_m = 30nm$. The light propagating through the PC is scattered into various diffraction orders (plane waves), which we labeled as $(n_1, n_2)$. Their field amplitudes, $R_{\sigma n_1 n_2}$, (with $\sigma = s$, and $\sigma = p$ staying for s- and p-polarizations, respectively), can be individually determined from a linear system of equations (see Supplementary Material, Section 1, Eq.77). In Figures 2b,f,g, we show the spectra of the normalized reflection coefficient, $\Delta R$ (solid black line), for three values of the twisting angle $\Phi$, as illustrated by the schematics in Figure 2a,e,i. Here $\Delta R = R_b - R$, with $R_b$ being the field amplitude of the $(0, 0)$ order mode of an α-MoO$_3$ layer on top of an unstructured gold film, and $R$ being the scattered amplitude of the $(0, 0)$ order mode in the twisted PC. At $\Phi = 0$ (Figure 2b), $\Delta R$ displays a peak at $885.1 cm^{-1}$. At the same frequency, the $R_{p10}$ amplitude (blue curve) shows a clear resonant peak in its spectrum, indicating the emergence of a PhP Bragg resonance in $(\pm 1, 0)$ diffraction order. Indeed, the latter largely exceeds the contribution from the other nearest diffraction orders $R_{p1-1}$ and $R_{p11}$, as follows from their spectra, shown by the red and yellow lines, respectively. Apart from the $(\pm 1, 0)$ order mode, other Fourier field harmonics can also display resonant peaks, such as at $881.4 cm^{-1}$, where the dominant amplitudes come from the $(\pm 1, \pm 1)$ and $(\pm 1, \mp 1)$ orders. In these pairs, the sign of the left number matches the sign of the right number (up for up, down for down).

The emerging Bragg resonances can be explained by matching the IFC of the M1 PhP mode in the structure without holes with the reciprocal space vectors $\boldsymbol{G_{n_1 n_2}} = n_1 \boldsymbol{g_1} + n_2 \boldsymbol{g_2}$, pointing from the origin towards the reciprocal space points (the grid of integers), marked by the blue dots in Figures 2d,h,l. For instance, at $885.1 cm^{-1}$, the IFC of the M1 PhP mode (indicated by the green dashed line in Figure 2d and also seen as the bright maxima of the colorplot – see Supplementary Material, Section 2 for more details about colorplot), meets the reciprocal vector $\boldsymbol{G_{10}}$, thus fulfilling the Bragg resonance condition for the $(\pm 1, 0)$ order. When the resonance condition is met, the electric field pattern, $E_z(x, y)$, represents a standing wave with the reciprocal vector $\boldsymbol{G_{10}}$, as shown in Figure 2c.

Next, we analyze the spectra of the PC when the α-MoO$_3$ layer is twisted by 20° and 40° with respect to the HA (schematics in Figure 2e,i), Figure 2f and 2j, respectively. In both cases, the frequency of the resonant peaks redshifts, to $880.8 cm^{-1}$ for $\Phi = 20°$ and to $868.2 cm^{-1}$ for $\Phi = 40°$, respectively, thus amounting up to ~11% of the entire RB. This twist-induced shift is also seen in the reciprocal space representations shown in Figure 2h,l. Due to the twist of the lattice in the real space, the reciprocal space points undergo an anticlockwise rotation. Because of the anisotropy of α-MoO$_3$, the dispersion relation depends upon the orientation of the k-vector and thus the dispersion curves. Consequently, the frequency at which the IFC meets the reciprocal vector $\boldsymbol{G_{10}}$ varies, so that the Bragg resonance condition is fulfilled at $880.8 cm^{-1}$ and $868.2 cm^{-1}$

for 20° and 40°, respectively. At these frequencies, the field patterns shown in Figure 2g,k clearly display a Bloch standing wave whose fringes are oriented along the $L_1$ direction. Consistently, the major contribution into the excited PhP near-field arises from the diffraction order $(\pm 1, 0)$ with the dominating field amplitude, $R_{p10}$. Note that the Bragg resonance condition for other diffraction orders also shifts in frequency, as observed in the peak position of the field amplitudes $R_{p11}$ and $R_{p1-1}$. Actually, for non-zero $\Phi$ their spectra no longer coincide, due to the break of symmetry. For instance, for $\Phi = 20°$, when the IFC meets $(1,1)$ and $(-1,-1)$ reciprocal space points, it does not meet those at $(1,-1)$ and $(-1,1)$. Finally, remark that the frequency tunability of the Bragg resonance by rotating the lattice with respect to the optical axes of the crystal layer remains an important intrinsic feature of twisted PCs[16], [17].

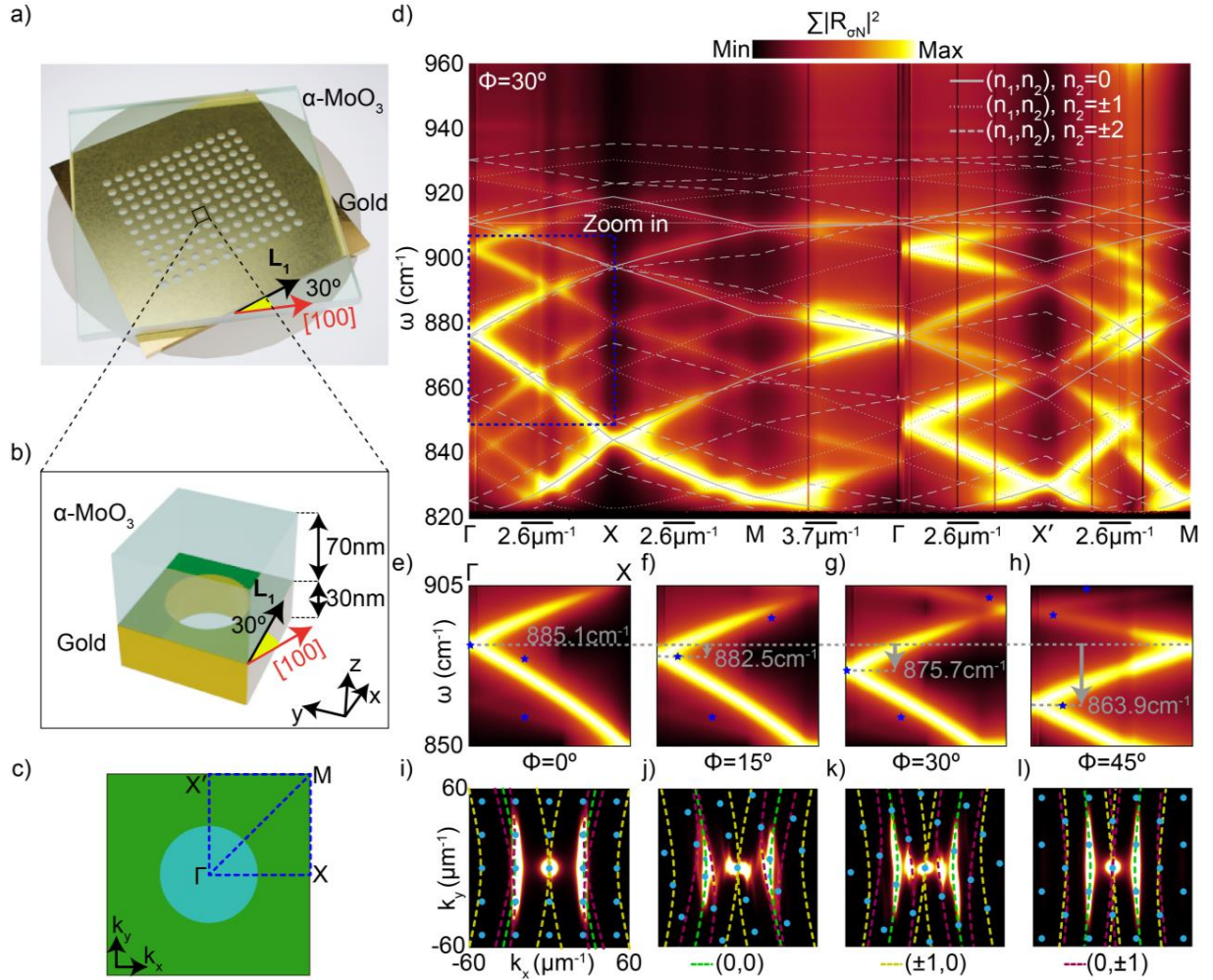

**Figure 3. Tunability of the band structure of the twisted PC.** (a) Schematic of the twisted PC for $\Phi = 30°$. (b) A zoom-in view of the Wigner-Seitz cell of the PC. The thickness of α-MoO₃ layer and the gold film at $d = 70nm$, and $d_m = 30nm$, respectively. (c) A sketch of the reciprocal space directions, indicated by blue dashed lines, with the main points of the BZ labeled as $\Gamma$, $X$, $M$ and $X'$, respectively. The BZ is combined with a schematic of the x-y projection of the Wigner-Seitz cell. (d) Color plot (representing $\sum |R_{\sigma N}|^2$ for $\sigma \in \{s, p\}$) indicates the band structure of the twisted PC for $\Phi = 30°$. The empty lattice band structure is traced by the gray lines, for the

diffraction orders $(n_1, 0)$, $(n_1, \pm 1)$ and $(n_1, \pm 2)$, with $n_1$ restricted to $-2 \leq n_1 \leq 2$. **(e, f, g, h)** Zoom-ins of the band structure for $\Phi$ of 0°, 15°, 30° and 45°, respectively. The shift of the bands at the $\Gamma$ point is indicated by the gray dashed lines and the arrow. The blue asterisk symbols indicate the extracted momentum from experiments (See details in Supplementary Material, Section 3). **(I, j, k, l)** Color plots indicating the IFCs of the twisted PC for $\Phi$ of 0°, 15°, 30° and 45°, at the frequencies of $885.1 cm^{-1}, 882.5 cm^{-1}, 875.7 cm^{-1}$, and $863.9 cm^{-1}$, respectively. The reciprocal lattice is shown by the blue dots, and the IFCs of the empty lattice for the diffraction orders $(0,0)$, $(\pm 1, 0)$ and $(0, \pm 1)$ are represented by the green, yellow and red dashed lines, respectively.

The dependence of the Bragg resonances upon the rotation angle is also encoded into the band structure of the twisted PC, which is illustrated in Figure 3. The volumetric Wigner-Seitz cell of the twisted PC is schematically shown in Figures 3a-c, together with the high symmetry points of the first Brillouin Zone (BZ). Apart from the square lattice symmetry points $\Gamma$, X and M, we have introduced an additional point, X′, as obviously due to the in-plane anisotropy of the α-MoO$_3$ layer, the PhP dispersion along [100] and the [010] axes is different. More specifically, the band structure along the path starting from $\Gamma$ and finishing in X differs from that connecting $\Gamma$ and X′. The color plot shown in Figure 3d illustrates the near-field intensity given by the summation of a large number of the field amplitudes, $R_{\sigma n_1 n_2}$, across the path $\Gamma\rightarrow$ X$\rightarrow$ M$\rightarrow$ $\Gamma\rightarrow$ X′$\rightarrow$ M (the arrows indicate the direction of the path, from the starting point to the end point) in the in-plane momentum plane. The bright maxima of the color plot approximately represent the "density of states" of PhP modes in the twisted PC. For a better interpretation of the band structure, we also plot in Figure 3d (gray dashed lines) the $\Phi$-dependent empty grating dispersion branches, $\boldsymbol{k}_{n_1 n_2}(\omega) = \boldsymbol{k}_{M1}(\omega) + \boldsymbol{G}_{n_1 n_2}$ (where $\boldsymbol{k}_{M1}$ is the momentum of PhPs supported by α-MoO$_3$ on top of an unstructured metal film), for the diffraction orders $-2 \leq n_1, n_2 \leq 2$ and for $\Phi = 30°$. The empty grating dispersion finds an excellent agreement with the maxima of the color plot. Note that, the band gaps of the PhPs in the twisted PC are visually indistinguishable in the color plot. We attribute it to the low refractive index contrast, and thus the small reflectivity of the PhPs, by the areas of the holes. On the other hand, the dispersion branches of the twisted PC can be shifted by the twist angle, as illustrated in Figure 3e-h, showing a zoomed-in area within the blue dashed rectangle in Figure 3d. For instance in the $\Gamma$ point, while at $\Phi = 0°$ the two branches $-\boldsymbol{k}_{M1}(\omega) + \boldsymbol{G}_{10}$ and $\boldsymbol{k}_{M1}(\omega) + \boldsymbol{G}_{-10}$ meet at $885.1 cm^{-1}$ (Figure 3e), by increasing $\Phi$ to 15°, 30°, and 45°, the crossing redshifts to $882.5 cm^{-1}$, $875.7 cm^{-1}$, and $863.9 cm^{-1}$, respectively (in Figure 3f-h the shift is indicated by the gray dashed lines). To connect the tunability of the band structure with the Bragg resonance condition, in Figures 3i-l we represent the IFCs at the crossing frequencies, $885.1 cm^{-1}$, $882.5 cm^{-1}$, $875.7 cm^{-1}$, and $863.9 cm^{-1}$, respectively. For the reference, the IFCs of the empty lattice PhPs with momenta $\pm\boldsymbol{k}_{M1}, \pm\boldsymbol{k}_{M1} + \boldsymbol{G}_{\pm 10}$ and $\pm\boldsymbol{k}_{M1} + \boldsymbol{G}_{0\pm 1}$ are also rendered (green, yellow and red dashed lines, respectively). For each $\Phi$, we observe that the IFC meets the reciprocal space point $(\pm 1, 0)$ so that the Bragg resonance condition is fulfilled, implying the color plot maxima at this frequency at the $\Gamma$ point observed in Figures 3e-h. Remark that, while our discussion has been centered on the zoomed-in region, these findings extend to the entire PhP band structure, so that all the bands undergo frequency shifts by twisting.

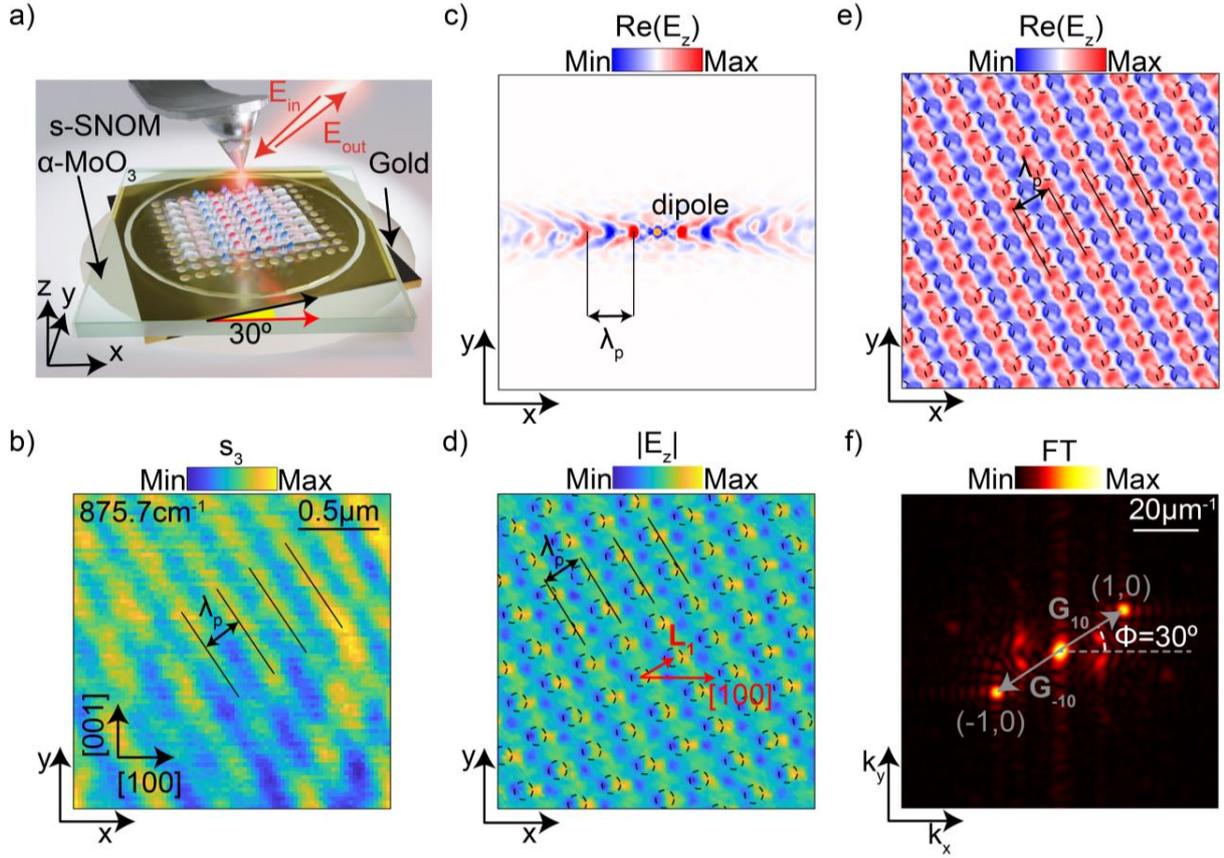

**Figure 4. Near-field imaging of the PhP modes in the twisted PC and their analysis. (a)** Schematics of the PC and the s-SNOM. The upper α-MoO₃ layer is twisted by $\Phi = 30°$ with respect to the HA in a thin gold layer. The structure is suspended in air, thus there is no substrate under the HA. An s-SNOM tip scans the PC under the presence of an incident field $E_{in}$, while the scattered field $E_{out}$ is recorded. **(b)** A near-field s-SNOM image taken at the third harmonic at $\omega = 875.5\ cm^{-1}$. Black lines indicate the fringes of the PhP Bloch wave. **(c)** The simulated $Re[E_z(x,y)]$ generated by a vertical point dipole on top of a finite-size twisted PC. The wavelength of the PhP along the *x*-direction is indicated by the black lines. **(d)** Simulation of the near-field image shown in **(b)**. The $L_1$ vector of the reciprocal lattice and the [100] crystallographic axis from α-MoO₃ are indicated by the red arrows, whereas the position of the holes are marked by the black dashed circles. Black solid lines indicate the fringes. **(e)** Simulated $Re[E_z(x,y)]$ generated by a normally incident plane wave polarized along [100] direction. Black lines indicate the fringes. **(f)** Colorplot depicting the FTs of the near-field image shown in **(b)**. The reciprocal space vectors $G_{10}$ and $G_{-10}$ are indicated by the gray arrows, forming an angle of 30° with respect to the $k_x$-axis.

To corroborate our theoretical analysis, we conducted experimental near-field measurements on twisted PCs, which, importantly, are fabricated using the same α-MoO₃ layer that is subsequently rotated on top of the HA in gold (see Methods). To visualize the PhPs Bloch modes, we employed near-field nanoimaging via scattering-type scanning near-field optical microscopy (s-SNOM)[23],

see schematics in Figure 4a. Both the sample and the s-SNOM tip are illuminated with a p-polarized mid-IR light at an incident frequency of $875.7 cm^{-1}$, i.e. within the RB frequency range. By recording the scattered field signal ($s_3$) (see Methods), we produced the near-field images visualizing PhPs excited in the twisted PC (color plot in Figure 4b). In the image we observe a series of parallel fringes, marked by black solid lines, with a separation distance between them matching the periodicity of the HA. To interpret the observed near-field pattern, we conducted full-wave simulations employing a vertical point dipole source, mimicking the s-SNOM tip[23]. The simulated snapshot of $Re(E_z)$ shown in Figure 4c reveals an oscillating field distribution (the wavelength is indicated by the black lines) within a narrow sector around the [100] crystallographic direction. This distribution is characteristic of recently reported canalized PhPs[9], [12]. The canalization regime of PhPs in our twisted PCs can be explained by the plane shape of the IFC of the M1 PhPs in an α-MoO₃ layer on top of a metal substrate, similar to α-MoO₃ on SiC[27]. Remarkably, such a directional pattern is produced by placing the dipole at any point above the twisted PC. Nevertheless, in the near-field image constructed by scanning the dipole in the x-y plane (Figure 4d, see Methods, near-field full-wave simulations section), one can clearly recognize a standing wave pattern, in good agreement with the near-field measurements (Figure 4b). Such periodic pattern clearly indicates a largely collective nature of the excited electromagnetic fields, inherent to PCs. Moreover, in both simulated and experimental near-field images, the direction perpendicular to the fringes (highlighted by the solid black lines) forms a $30°$ angle with the [100] crystallographic axis, thus being aligned with the lattice vector. This observation provides a hint that the $(\pm 1, 0)$ Bloch mode is excited due to the $(\pm 1, 0)$ Bragg resonance discussed above. To confirm our speculation, we performed full-wave simulations considering a normally-incident plane wave illumination, thus completely excluding any effects induced by a localized source. The resulting distribution of $Re(E_z)$ is illustrated in Figure 4e. The normally-incident wave "acquires" the momentum $\boldsymbol{G_{\pm 10}}$ provided by the lattice, so that the two excited contra-propagating PhPs plane waves form the $(\pm 1, 0)$ Bloch mode, which can be recognized in Figure 4e. Its field distribution agrees well in shape, wavelength, and fringe orientation with the near-field images in Figures 4b,d. Furthermore, by representing the Fourier transform (FT) of the experimental near-field image as a color plot in Figure 4f, we can observe two bright maxima labeled as $(1, 0)$ and $(-1, 0)$ that perfectly match the lattice vectors, $\boldsymbol{G_{\pm 10}}$. Overall, our simulations are consistent with our assumption that PhP Bloch mode is observed in the nanoimaging experiments.

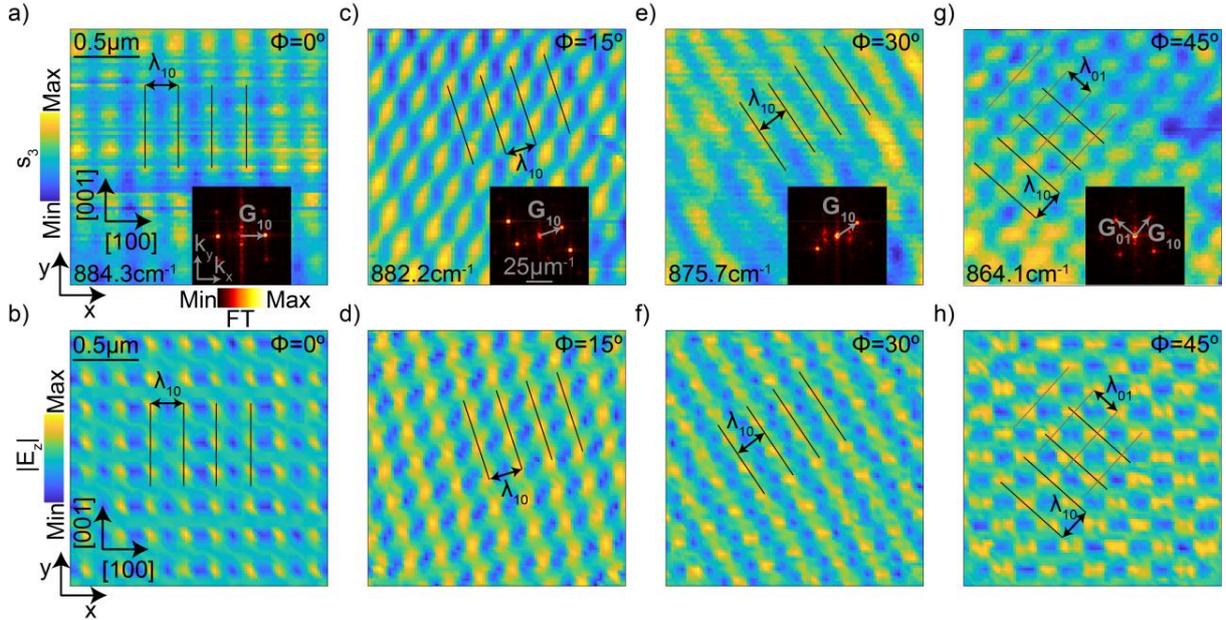

**Figure 5. Near-field imaging of the twist-tunable PC for different twist angles. (a, b, c, d)** s-SNOM near-field images for twist angles of $\Phi = 0°$, $15°$, $30°$ and $45°$ at the frequency of $884.3 cm^{-1}$, $882.2 cm^{-1}$, $875.7 cm^{-1}$, and $864.1 cm^{-1}$, respectively. Colorplots representing the FTs of the near-field images for each $\Phi$ are included as insets, where the reciprocal lattice vectors, $\boldsymbol{G_{10}}$ and $\boldsymbol{G_{01}}$, are indicated by the gray arrows. **(e, f, g, h)** Simulated near-field images for a finite-size twisted PC, for the $\Phi$ and $\omega$ as in **(a, c, e, g)**. In **(a, c, e, g)** the black solid lines indicate the fringes of the Bloch PhP wave in $(\pm 1,0)$ diffraction orders, while in **(g, h)** the gray lines indicate those in $(0, \pm 1)$ diffraction orders.

Our PhP nanoimaging experiments can be extrapolated to other twist angles, $\Phi$, as illustrated in Figure 5a-d. From panels a to d, α-MoO₃ layer maintains its crystallographic axes [100] and [001] aligned with the x and y axes, respectively whereas the HA is twisted anticlockwise. In Figure 5a-d, the near-field images are shown at frequencies $884.3 cm^{-1}$, $882.2 cm^{-1}$, $875.7 cm^{-1}$, and $864.1 cm^{-1}$, corresponding to $(\pm 1, 0)$ Bragg resonance condition for $\Phi = 0°$, $15°$, $30°$, and $45°$, respectively (near-field images of PhP Bloch modes in other diffraction orders $(n_1, n_2)$ apart from $(\pm 1, 0)$ are shown in Supplementary Material, Section 4). For each twisting angle, a standing wave with fringes perpendicular to the $\boldsymbol{L_1}$ lattice vector is observed, where the fringe positions are indicated by black solid lines. These measurements are in good agreement with the full-wave simulations conducted via the scanning dipole method, shown in Figures 5e-h. To analyze the PhP wavelength and the orientation of the fringes, we performed FTs of the near-field data, represented as color plots in the insets of Figures 5a-d. The positions of the FT maxima confirm that the reciprocal lattice vector, $\boldsymbol{G_{10}}$, is twisted by an angle $\Phi$ with respect to the $x$ axis. Specifically, for $\Phi = 45°$ in the near-field image (Figure 5d,h), in addition to the standing wave with fringes oriented along the lattice vector $\boldsymbol{L_1}$, we observe another standing wave oriented along $\boldsymbol{L_2}$ lattice vector (the fringes are indicated by gray lines). The appearance of the second standing wave arises due to the lattice symmetry. Indeed, for smaller twist angles, the Bragg resonance condition was fulfilled for two reciprocal lattice vectors, $\boldsymbol{G_{\pm 10}}$, since the IFC of the α-MoO₃ simultaneously

intersects with only two reciprocal space points, namely $(\pm 1, 0)$. However, for $\Phi = 45°$, the crystallographic axes of α-MoO₃ are positioned such that the intersection of the IFC with four reciprocal space points, $(\pm 1, 0)$ and $(0, \pm 1)$, becomes possible. This effect becomes even clearer in the Fourier transform of the measured data shown in the inset of Figure 5d, where four bright maxima match with reciprocal lattice vectors $\boldsymbol{G_{10}}$ and $\boldsymbol{G_{01}}$. Using our near-field images, it is also possible to reconstruct the momentum of the PhP Bloch modes for each angle Φ by measuring the separation distance between the fringes, $\lambda_{exp}$ (details provided in Supplementary Material, Section 3). We then translate the measured wavelength into momentum space ($G_{exp} = 2\pi/\lambda_{exp}$), which falls into the second BZ. To properly represent this momentum in the band structure, we fold this momentum from the second BZ to the first BZ by subtracting $g_1 - G_{exp}$. This folded momentum is depicted by blue asterisk symbols in Figure 3e-h, in good agreement with the theoretical predictions. Overall, in Figure 5 we have experimentally demonstrated tunability of Bloch modes by the twist angle.

To summarize, we have introduced and experimentally realized a low-loss tunable PC based on a heterostructure consisting of a twistable α-MoO₃ crystal layer on top of a gold HA. In a large contrast to previously reported twisted PCs, our design on the one hand preserves the pristine α-MoO₃ layer, and on the other hand, allows one to rotate it without any need to modify the lattice below it. While by the theoretical analysis we have demonstrated the twist-tuneability of the emerging Bragg resonances in our PC, our experiments clearly demonstrate the tunability of the PC band structure and the polaritonic Bloch modes. Remarkably, collective lattice effects largely dominate the natural polaritonic canalization effect. The latter is due to the intrinsically flat isofrequency curve of polaritons in a crystal layer above a metal substrate. Importantly, our general concept is valid for any similar heterostructure based on a biaxial crystal layer supporting polaritons. Our findings expand the vision of twistoptics in a wide perspective and particularly hold promise for the development of actively-rotatable two-dimensional polaritonic elements.

### 3. Methods

**Fabrication of the samples.** The gold HA samples were provided by the company NanoStruct GmbH. The sample consists of a gold film with a circular hole of $100 \mu m$ in diameter, placed on top of pristine glass. Simultaneously, a gold single crystal is fabricated via wet chemical synthesis[28], and subsequently transferred on top of the hole in the gold/glass substrate, ensuring a small physical overlap between the gold single crystal and the gold film. This setup allows the gold single crystal to present a gold/glass interface while being easily connected electrically to the sample holder via the gold film. The HAs are structured with focused ion beam milling (with a Ga source from a Zeiss Orion NanoFab), employing a $30 kV$ and $30 pA$ current, and a dwell time of $1 \mu s$ with 20 repeats. Furthermore, layers of α-MoO₃ with a thickness of about $70\ nm$ were exfoliated on top of the gold HAs through the application of a dry-transfer process[29]. The process began with a mechanical exfoliation of α -MoO₃ from commercially available bulk crystals sourced from Alfa Aesar, using Nitto tape (Nitto Denko, SPV 224P). Then, the α -MoO₃ flakes were transferred from the tape to a transparent stamp made of poly-(dimethylsiloxane) (PDMS).

Only the uniform flakes were selected using an optical microscope examination. Finally, a micromanipulator was used to precisely position the PDMS stamp with the α-MoO₃ flake onto the targeted area on the substrate before smoothly detaching the flake (dry-transfer technique).

**Near-field measurements.** The near-field optical studies were conducted using a state-of-the-art scattering-type scanning near-field optical microscope (s-SNOM, Neaspec GmbH), equipped with quantum cascade lasers from Daylight Solutions, which covered a spectral range of 850 to 1140 $cm^{-1}$. The s-SNOM combines an atomic force microscope (AFM) that operates in tapping mode with a frequency of approximately 285 $kHz$, a tapping amplitude near 200 $nm$, and commercially available metal-coated (Pt/Ir) AFM tips (ARROW-NCPt-50, Nanoworld). The process involved directing p-polarized mid-infrared light at the sample surface and focusing the light scattered back by the tip onto an infrared detector (Komar Technologies) using a parabolic mirror. We employed a pseudoheterodyne interferometric approach to decode the amplitude and phase of the signal, and the signal was demodulated at the third harmonic (the signal amplitude is denoted as $s_3$) to minimize the influence of far-field background scattering.

**Mode expansion calculations.** We develop an analytical theory based on Maxwell's equations. We represented electromagnetic fields as quasi-eigenmodes in our four-region structure. The four regions are: (1) an isotropic superstrate characterized by permittivity $\varepsilon_1$ (we considered air, $\varepsilon_1 = 1$); (2) a biaxial layer (α-MoO₃) with a $3 \times 3$ permittivity tensor $\hat{\varepsilon}$ and thickness $d$; (3) the gold HA (metal film with the thickness $d_m$, lattice vectors $\boldsymbol{L_1}$ and $\boldsymbol{L_2}$, and hole radius $a$, where the holes are filled by air, thus $\varepsilon_h = 1$.); and (4), the isotropic substrate, with the dielectric permittivity $\varepsilon_4$. The theoretical framework is based on two key assumptions: (i) the use of the quasi-orthogonal basis vectors to describe wave propagation within an infinite continuous anisotropic slab[30], and (ii) the approximation of the metal film by a perfect electric conductor – which is common and reasonable in the considered frequency range – so that the electromagnetic fields can only exist inside the holes[31]. More details are provided in the Supplementary Material, Section 1. The resulting dispersion relation implies an infinite sum over all the spatial field harmonics. This set of harmonics was truncated to $max(|n_1|) = max(|n_2|) = 12$.

**Far-field full-wave simulations.** Full-wave simulations of infinite PC illuminated by a normal plane wave, based on the finite-element method in the frequency domain, were performed using COMSOL. We considered an infinite two-dimensional square periodic gold HA with the thickness $d_m = 30nm$, period $L = 250nm$, and the hole radius $a = 55nm$, where the holes are filled by air ($\varepsilon_h = 1$). On top, we place a α-MoO₃ layer with thickness $d = 70nm$, whose crystallographic axes are twisted $-30°$ respect to the x-y plane. Both the superstrate and substrate are set as air, thus $\varepsilon_1 = \varepsilon_4 = 1$. A normal incident plane wave linearly polarized along the [100] crystallographic axis.

**Near-field full-wave simulations.** For the scanning point dipole simulations, we employed full-wave simulations using COMSOL. We modeled a finite $10 \times 10$ two-dimensional square periodic gold HA with thickness $d_m = 30nm$, period $L = 250nm$, and the hole radius $a = 55nm$, where the holes are filled by air ($\varepsilon_h = 1$). The lattice vector $\boldsymbol{L_1}$ is twisted at a certain positive angle $\Phi$ with respect to [100] crystallographic axis, and the entire HA is surrounded by an air crown gap with inner diameter of $3.5\mu m$ and outer diameter of $4\mu m$, mimicking the experimental setup,

illustrated in Figure 4a. On top, we place a α-MoO$_3$ layer with thickness $d = 70nm$, with crystallographic axis [100] located in the x axis. Both the superstrate and substrate are set as air, thus $\varepsilon_1 = \varepsilon_4 = 1$. A vertically oriented point dipole is placed at $300nm$ distance from the surface, while $|E_z|$ field component is recorded at $20nm$ from the surface, at the same x-y position as the dipole. The dipole is scanned in the x-y plane.

## Acknowledgments

# References

[1] D. N. Basov, M. M. Fogler, and F. J. G. de Abajo, "Polaritons in van der Waals materials," *Science*, vol. 354, no. 6309. p. aag1992, 2016.
[2] T. Low *et al.*, "Polaritons in layered two-dimensional materials," *Nature Materials*, vol. 16, no. 2, pp. 182–194, Feb. 2017, doi: 10.1038/nmat4792.
[3] Q. Zhang *et al.*, "Interface nano-optics with van der Waals polaritons," *Nature*, vol. 597, no. 7875, pp. 187–195, Sep. 2021, doi: 10.1038/s41586-021-03581-5.
[4] W. Ma *et al.*, "Ghost hyperbolic surface polaritons in bulk anisotropic crystals," *Nature*, vol. 596, no. 7872, pp. 362–366, Aug. 2021, doi: 10.1038/s41586-021-03755-1.
[5] W. Ma *et al.*, "In-plane anisotropic and ultra-low-loss polaritons in a natural van der Waals crystal," *Nature*, vol. 562, no. 7728, pp. 557–562, Oct. 2018, doi: 10.1038/s41586-018-0618-9.
[6] J. Taboada-Gutiérrez *et al.*, "Broad spectral tuning of ultra-low-loss polaritons in a van der Waals crystal by intercalation," *Nature Materials*, vol. 19, no. 9, pp. 964–968, Sep. 2020, doi: 10.1038/s41563-020-0665-0.
[7] N. C. Passler *et al.*, "Hyperbolic shear polaritons in low-symmetry crystals," *Nature*, vol. 602, no. 7898, pp. 595–600, Feb. 2022, doi: 10.1038/s41586-021-04328-y.
[8] G. Hu *et al.*, "Topological polaritons and photonic magic angles in twisted α-MoO3 bilayers," *Nature*, vol. 582, no. 7811, pp. 209–213, Jun. 2020, doi: 10.1038/s41586-020-2359-9.
[9] J. Duan *et al.*, "Twisted Nano-Optics: Manipulating Light at the Nanoscale with Twisted Phonon Polaritonic Slabs," *Nano Lett.*, vol. 20, no. 7, pp. 5323–5329, Jul. 2020, doi: 10.1021/acs.nanolett.0c01673.
[10] M. Chen *et al.*, "Configurable phonon polaritons in twisted α-MoO3," *Nature Materials*, vol. 19, no. 12, pp. 1307–1311, Dec. 2020, doi: 10.1038/s41563-020-0732-6.
[11] Z. Zheng *et al.*, "Phonon Polaritons in Twisted Double-Layers of Hyperbolic van der Waals Crystals," *Nano Lett.*, vol. 20, no. 7, pp. 5301–5308, Jul. 2020, doi: 10.1021/acs.nanolett.0c01627.
[12] J. Duan *et al.*, "Multiple and spectrally robust photonic magic angles in reconfigurable α-MoO3 trilayers," *Nature Materials*, vol. 22, no. 7, pp. 867–872, Jul. 2023, doi: 10.1038/s41563-023-01582-5.
[13] C. Zheng, G. Hu, X. Liu, X. Kong, L. Wang, and C.-W. Qiu, "Molding Broadband Dispersion in Twisted Trilayer Hyperbolic Polaritonic Surfaces," *ACS Nano*, vol. 16, no. 8, pp. 13241–13250, Aug. 2022, doi: 10.1021/acsnano.2c07123.
[14] O. G. Matveeva *et al.*, "Twist-tunable polaritonic nanoresonators in a van der Waals crystal," *npj 2D Materials and Applications*, vol. 7, no. 1, p. 31, Apr. 2023, doi: 10.1038/s41699-023-00387-z.
[15] S. Dai *et al.*, "Tunable Phonon Polaritons in Atomically Thin van der Waals Crystals of Boron Nitride," *Science*, vol. 343, no. 6175. pp. 1125–1129, 2014.
[16] N. Capote-Robayna, O. G. Matveeva, V. S. Volkov, P. Alonso-González, and A. Y. Nikitin, "Twisted Polaritonic Crystals in Thin van der Waals Slabs," *Laser & Photonics Reviews*, vol. 16, no. 9, p. 2200428, Sep. 2022, doi: 10.1002/lpor.202200428.
[17] J. Lv *et al.*, "Hyperbolic polaritonic crystals with configurable low-symmetry Bloch modes," *Nature Communications*, vol. 14, no. 1, p. 3894, Jul. 2023, doi: 10.1038/s41467-023-39543-w.
[18] N. R. Sahoo *et al.*, "Polaritons in Photonic Hypercrystals of van der Waals Materials," *Advanced Functional Materials*, vol. n/a, no. n/a, p. 2316863, Jul. 2024, doi: 10.1002/adfm.202316863.
[19] A. Yu. Nikitin, F. Guinea, and L. Martin-Moreno, "Resonant plasmonic effects in periodic graphene antidot arrays," *Applied Physics Letters*, vol. 101, no. 15, p. 151119, Oct. 2012, doi: 10.1063/1.4760230.
[20] L. Xiong *et al.*, "Photonic crystal for graphene plasmons," *Nature Communications*, vol. 10, no. 1, p. 4780, Oct. 2019, doi: 10.1038/s41467-019-12778-2.

[21] S. S. Sunku *et al.*, "Photonic crystals for nano-light in moiré graphene superlattices," *Science*, vol. 362, no. 6419, pp. 1153–1156, Dec. 2018, doi: 10.1126/science.aau5144.
[22] F. J. Alfaro-Mozaz *et al.*, "Deeply subwavelength phonon-polaritonic crystal made of a van der Waals material," *Nature Communications*, vol. 10, no. 1, p. 42, Jan. 2019, doi: 10.1038/s41467-018-07795-6.
[23] F. J. Alfaro-Mozaz *et al.*, "Hyperspectral Nanoimaging of van der Waals Polaritonic Crystals," *Nano Lett.*, vol. 21, no. 17, pp. 7109–7115, Sep. 2021, doi: 10.1021/acs.nanolett.1c01452.
[24] J. Yang *et al.*, "Near-Field Excited Archimedean-like Tiling Patterns in Phonon-Polaritonic Crystals," *ACS Nano*, vol. 15, no. 5, pp. 9134–9142, May 2021, doi: 10.1021/acsnano.1c02507.
[25] S. Guddala *et al.*, "Topological phonon-polariton funneling in midinfrared metasurfaces," *Science*, vol. 374, no. 6564. pp. 225–227, 2021.
[26] G. Álvarez-Pérez *et al.*, "Infrared Permittivity of the Biaxial van der Waals Semiconductor α-MoO3 from Near- and Far-Field Correlative Studies," *Advanced Materials*, vol. 32, no. 29, p. 1908176, Jul. 2020, doi: 10.1002/adma.201908176.
[27] J. Duan *et al.*, "Canalization-based super-resolution imaging using a single van der Waals layer," *arXiv:2404.14876*, 2024.
[28] E. Krauss *et al.*, "Controlled Growth of High-Aspect-Ratio Single-Crystalline Gold Platelets," *Crystal Growth & Design*, vol. 18, no. 3, pp. 1297–1302, Mar. 2018, doi: 10.1021/acs.cgd.7b00849.
[29] A. Castellanos-Gomez *et al.*, "Deterministic transfer of two-dimensional materials by all-dry viscoelastic stamping," *2D Materials*, vol. 1, no. 1, p. 011002, Apr. 2014, doi: 10.1088/2053-1583/1/1/011002.
[30] G. Álvarez-Pérez, K. V. Voronin, V. S. Volkov, P. Alonso-González, and A. Y. Nikitin, "Analytical approximations for the dispersion of electromagnetic modes in slabs of biaxial crystals," *Phys. Rev. B*, vol. 100, no. 23, p. 235408, Dec. 2019, doi: 10.1103/PhysRevB.100.235408.
[31] A. Roberts, "Electromagnetic theory of diffraction by a circular aperture in a thick, perfectly conducting screen," *J. Opt. Soc. Am. A*, vol. 4, no. 10, pp. 1970–1983, Oct. 1987, doi: 10.1364/JOSAA.4.001970.



\newpage 

\title{\Large{\textbf{Supplementary Material for \\ \vspace{0.5 cm} Low-loss twist-tunable in-plane anisotropic polaritonic crystals}}}

\vspace{8 pt}
\setstretch{0.1}
\author[1,2]{\normalsize{Nathaniel Capote-Robayna}}
\author[3,6]{\normalsize{Ana I.F. Tresguerres-Mata}}
\author[3,6]{\normalsize{Aitana Tarazaga Mart\'in-Luengo}}
\author[3,6]{\normalsize{Enrique Ter\'an-Garc\'ia}}
\author[4,5]{\normalsize{Luis Martin-Moreno}}
\author[3,6*]{\normalsize{Pablo Alonso-Gonz\'alez}}
\author[2,7*]{\normalsize{Alexey Y. Nikitin}}
\affil[1]{\small{\textit{Applied Physics department, Engineering school of Gipuzkoa, University of the Basque Country (UPV/EHU), Donostia-San Sebasti\'an, 20018, Spain.}}}
\affil[2]{\small{\textit{Donostia International Physics Center (DIPC), Donostia-San Sebasti\' an, 20018, Spain.}}}
\affil[3]{\small{\textit{Department of Physics, University of Oviedo, Oviedo, 30006, Spain.}}}
\affil[4]{\small{\textit{Instituto de Nanociencia y Materiales de Arag\'on (INMA), CSIC-Universidad de Zaragoza, Zaragoza, 50009, Spain.}}}
\affil[5]{\small{\textit{Departamento de Física de la Materia Condensada, Universidad de Zaragoza, Zaragoza, 50009, Spain.}}}
\affil[6]{\small{\textit{Center of Research on Nanomaterials and Nanotechnology, CINN (CSIC-Universidad de Oviedo), El Entrego, 33940, Spain.}}}
\affil[7]{\small{\textit{IKERBASQUE, Basque Foundation for Science, Bilbao, 48013, Spain.}}}
\affil[*]{\small{\textit{Corresponding author. Email: \href{mailto:pabloalonso@uniovi.es}{pabloalonso@uniovi.es}, \href{mailto:alexey@dipc.org}{alexey@dipc.org}}}}
\date{}
{\let\clearpage\relax%
\maketitle }

\newpage

\tableofcontents

\normalsize
\setstretch{0.5}

\section{Derivation of PhP dispersion relation in the twisted PC}

In this section, we will derive the linear system of equations for the amplitudes of Fourier harmonics of scattered electric fields shown in the main text, following closely the mode expansion used in studies of extraordinary optical transmission~\cite{Martin2010}. This system will be needed to analyze the dispersion of polaritonic modes in our polaritonic crystal. The crystal is composed of four layers arranged in the following sequence from bottom to top: a half-infinite substrate, hole array in a gold film, an anisotropic slab, and a half-infinite air superstrate. To embark on the mathematical derivation, we will employ the basis for ordinary and extraordinary modes within the anisotropic material and s- and p- polarized waves in the regions filled with the isotropic dielectric. Furthermore, we assume that the metallic slab behaves as a perfect electric conductor, thus wave propagation is allowed exclusively inside the holes, where the fields will be represented in the form of the waveguiding modes.

\begin{figure}[h]
    \centering
    \includegraphics[scale=0.55]{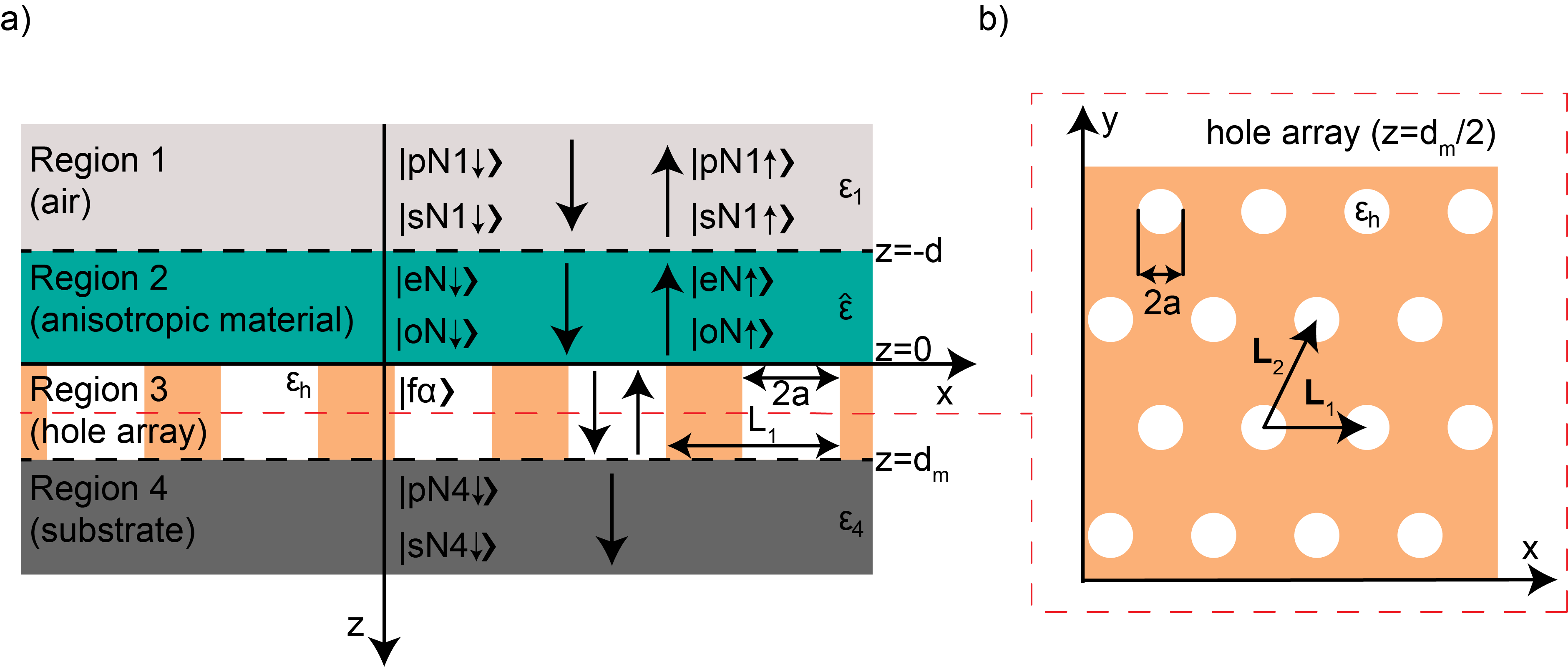}
    \caption{(a) Schematic depicting the four regions and the basis vectors used to describe the electromagnetic fields in the whole structure. The first medium comprises a semi-infinite layer of air with permittivity $\varepsilon_1$, extending from $-\infty<z<-d$. The second medium consists of a slab of an anisotropic material with the dielectric permittivity tensor $\hat{\varepsilon}$, spanning $-d<z<0$. The third medium comprises a metallic layer with a periodic hole array with periodicity $L_1$ through the x direction and radius $a$, where an isotropic material is filling the holes with the dielectric permittivity $\varepsilon_h$, spanning $0<z<d_{m}$. The fourth medium consists of a semi-infinite layer of substrate with permittivity $\varepsilon_4$, spanning $d_{m}<z<\infty$. The $z$-axis is aligned with the direction of incident plane wave propagation. (b) In-plane view from the periodic hole array in the metallic slab at the position $z=d_{m}/2$. The holes have radius $a$ and are filled by isotropic material with permittivity $\varepsilon_h$, distributed in a 2D lattice defined by vectors $\textbf{L}_1$ and $\textbf{L}_2$.}
    \label{fig:4_regions}
\end{figure}

To be more specific, we divide the space into four regions, as illustrated in Figure~\ref{fig:4_regions}. Region 1 represents the incidence medium, assumed to be isotropic with the dielectric permittivity $\varepsilon_1$, and spans the range $-\infty < z < -d$. Region 2 encompasses the anisotropic material, for instance $\alpha\textrm{-MoO}_3$, confined within $-d \leq z \leq 0$. Region 3 hosts the metallic layer featuring periodic circular holes with radius $a$ (we assume the holes to be filled by an isotropic material with the dielectric permittivity $\varepsilon_h$), situated within $0 < z < d_{m}$. The periodicity is characterized by the in-plane basis vectors $\textbf{L}_1=(L_{1x},L_{1y})$ and $\textbf{L}_2=(L_{2x},L_{2y})$, corresponding to the reciprocal lattice vectors defined as $\textbf{g}_1=(g_{1x},g_{1y})$ and $\textbf{g}_2=(g_{2x},g_{2y})$, satisfying the condition $\textbf{L}_i\textbf{g}_j=2\pi \delta_{ij}$. Finally, Region 4 accommodates an isotropic substrate with the dielectric permittivity $\varepsilon_4$, spanning $d_{m} \leq z < \infty$. Note that the orientation of the z-axis points in the same direction as the incident plane wave propagation.

As for the anisotropic material within the second region, we assume that its dielectric permittivity can be described by a diagonal $3\times 3$ tensor in the coordinate system aligned with crystallographic axes:

\begin{equation}
    \widehat{\varepsilon}=\begin{pmatrix}
    \varepsilon_x & 0 & 0 \\
    0 & \varepsilon_y & 0 \\
    0 & 0 & \varepsilon_z
    \end{pmatrix}.
\end{equation}

We will use different vector basis in each medium. Firstly, to represent the electric fields in isotropic regions 1 and 4, we introduce the following basis in Dirac notations:

\begin{equation}
\label{eq:sp basis}
    \langle \textbf{r}|sN\xi\uparrow,\downarrow\rangle_{3D}=\frac{1}{q_N}\begin{pmatrix}
    -q_{yN}\\
    q_{xN}\\
    0
    \end{pmatrix}e^{i\textbf{k}_N\textbf{r}},\quad  \langle \textbf{r}|pN\xi\uparrow,\downarrow\rangle_{3D}=\frac{1}{q_N}\begin{pmatrix}
    q_{xN}\\
    q_{yN}\\
    \pm\frac{q_N^2}{q_{\xi zN}}
    \end{pmatrix}e^{i\textbf{k}_N\textbf{r}},
\end{equation}
were $\xi=\{1,4\}$ labels the respective media, labels ''s ''and ''p'' relate to s- and p-polarizations, respectively, and the arrow indicates the propagation direction through the media, denoting $+$($-$) for $\uparrow$ ($\downarrow$), respectively (the corresponding direction of propagation of the electromagnetic waves is depicted by arrows in Figure~\ref{fig:4_regions}). Here, $q_{xN}$ and $q_{yN}$ denotes the x- and y- components of the normalized in-plane momentum, $\textbf{q}_N$,  while $q_{\xi zN}$ is its out of plane (z-) component. Their analytical expressions read as follows:

\begin{equation}
    q_{xN}=\frac{k_{xN}}{k_0},\quad k_{xN}=k_{x}+n_1g_{1x}+n_2g_{2x},
\end{equation}

\begin{equation}
    q_{yN}=\frac{k_{yN}}{k_0},\quad k_{yN}=k_{y}+n_1g_{1y}+n_2g_{2y},
\end{equation}

\begin{equation}
    \textbf{q}_N=\frac{\textbf{k}_N}{k_0},\quad \textbf{k}_N=(k_{xN},k_{yN},0),
\end{equation}

\begin{equation}
    q_{\xi zN}=\frac{k_{\xi zN}}{k_0}, \quad  k_{\xi zN}=\sqrt{\varepsilon_{\xi}k_0^2-k_N^2},
\end{equation}
where $k_x$ and $k_y$ represent the x- and y- components of the in-plane incident momentum, and $k_0=2\pi /\lambda$, with $\lambda$ being the incident wavelength, and N is a multi-index containing information about both index $n_1$ and $n_2$ ($N\equiv\{n_1,n_2\}$). It is worth mentioning that the reciprocal lattice is represented with the help of the vectors $n_1\textbf{g}_1+n_2\textbf{g}_2$, for any $\{n_1,n_2\}\in\mathbb{N}$, so that $\textbf{k}_N$ is the vectorial momentum of the mode $(n_1,n_2)$. Throughout the derivation, we will use bold letters for vectors and non-bold for their norms, for instance $k_N=\sqrt{k_{xN}^2+k_{yN}^2} $. 

In the representation of the fields inside the anisotropic slab, the choice of the basis depends on the momenta of ordinary and extraordinary waves~\cite{AlvarezPerez19}. The out-of-plane component of their normalized momentum is expressed as below:

\begin{equation}
\label{eq:qz_ord_extr}
    q^2_{\gamma zN}=-\bigg[\frac{1}{2}\bigg(\frac{\varepsilon_x+\varepsilon_z}{\varepsilon_z}q_{xN}^2+\frac{\varepsilon_y+\varepsilon_z}{\varepsilon_z}q_{yN}^2-(\varepsilon_x+\varepsilon_y)\bigg)\pm\frac{1}{2}\sqrt{Q_N}\bigg],
\end{equation}
where $\gamma=\{o,e\}$, the sign + (-) correspond to the ordinary (extraordinary) mode and $Q_N$ denotes:

\begin{equation}
    Q_N=\bigg(\varepsilon_x-\varepsilon_y+\frac{\varepsilon_z-\varepsilon_x}{\varepsilon_z}q_{xN}^2-\frac{\varepsilon_z-\varepsilon_y}{\varepsilon_z}q_{yN}^2\bigg)^2+4\frac{(\varepsilon_z-\varepsilon_x)(\varepsilon_z-\varepsilon_y)}{\varepsilon^2_z}q^2_{xN}q^2_{yN}.
\end{equation}
We will use the $k_{\gamma zN}=k_0q_{\gamma zN}$ for the non-normalized out-of-plane momentum to simplify the notations, where necessary. The basis for ordinary and extraordinary waves can be written as~\cite{AlvarezPerez19}:

\begin{equation}
\label{eq:oe basis}
    \langle \textbf{r}|oN\uparrow,\downarrow\rangle_{3D}=\frac{1}{\chi^{o}_Nq_N}\begin{pmatrix}
    -q_{yN}(1-\Delta_{1N}\Delta_{zN})\\
    q_{xN}\\
    \pm q_{xN}q_{yN}q_{ozN}\Delta_{1N}
    \end{pmatrix}e^{i\textbf{k}_N\textbf{r}},\quad \langle \textbf{r}|eN\uparrow,\downarrow\rangle_{3D}=\frac{1}{\chi^{e}_Nq_N}\begin{pmatrix}
    q_{xN}\frac{\Delta_{2N}-q^2_{yN}}{\Delta^e_{xN}}\\
    q_{yN}\\
    \pm\frac{\Delta_{2N}}{q_{ezN}}
    \end{pmatrix}e^{i\textbf{k}_N\textbf{r}},
\end{equation}
where the $+$ ($-$) sign corresponds to $\uparrow$ ($\downarrow$) propagation direction and the normalization factors $\chi^{o}_N$ and $\chi^{e}_N$ are expressed as:

\begin{equation}
    \chi^{o}_N=\sqrt{1+\frac{q_{yN}^2\Delta_{1N}\Delta_{zN}}{q^2_{N}}\bigg(\Delta_{1N}\Delta_{zN}-2\bigg)}, \quad \chi^{e}_N=\sqrt{1+\frac{q^2_{xN}}{q^2_{N}}\bigg(\frac{\Delta_{2N}-q^2_{yN}}{\Delta^e_{xN}}\bigg)^2-\frac{q^2_{xN}}{q^2_{N}}},
\end{equation}
and the auxiliary functions $\Delta$ have the following expressions:
\begin{equation}
    \Delta_{xN}^{\gamma}=\varepsilon_{x}-q^2_{yN}-q_{\gamma zN}^2, \quad \Delta_{yN}^{\gamma}=\varepsilon_{y}-q^2_{xN}-q_{\gamma zN}^2, \quad 
    \Delta_{zN}=\varepsilon_{z}-q^2_{xN}-q_{yN}^2,
\end{equation}
\begin{equation}
    \Delta_{1N}=\frac{\Delta^o_{xN}-q_{xN}^2}{\Delta_{zN}\Delta^o_{xN}-q_{xN}^2q_{ozN}^2},\quad \Delta_{2N}=\frac{\Delta^e_{xN}\Delta^e_{yN}-q_{xN}^2q_{yN}^2}{\Delta_{xN}^e-q_{xN}^2},
\end{equation}
for $\gamma\in\{o,e\}$.

Within region 3, we consider the metal to be a perfect electric conductor, resulting in analytical field expressions inside the holes. We can neglect transverse magnetic modes (TM) and focus solely to transverse electrical (TE) modes~\cite{Roberts87}. The electric field for the TE modes in circular holes can be explicitly written in cylindrical coordinates $\{r,\theta,z\}$ as: 

\begin{equation}
    E_{r,nml}=\frac{na}{u_{nm}r}J_n\left(u_{nm}\frac{r}{a}\right)C_{r,nl}(\theta)A_{nm}e^{\pm iz\nu_{nm}},
\end{equation}
\begin{equation}
    E_{\theta,nml}=-J_n'\left(u_{nm}\frac{r}{a}\right)C_{r,nl}(\theta)A_{nm}e^{\pm iz\nu_{nm}},
\end{equation}
\begin{equation}
    E_{z,nml}=0,
\end{equation}
where $J_{n}$ is the n-th Bessel function of the 1-st kind, 

\begin{equation}
    \nu_{nm}=\sqrt{\varepsilon_hk_0^2-\frac{u^2_{nm}}{a^2}},\quad A_{nm}=\sqrt{\frac{2-\delta_{n0}}{\pi}}\frac{u_{nm}}{a}\frac{1}{J_n(u_{nm})\sqrt{u_{nm}^2-n^2}},
\end{equation}
and

\begin{equation}
    C_{r,nl}(\theta)=\bigg\{
    \begin{aligned}
        & \cos n\theta, \quad l=horizontal,\\
        & -\sin n\theta, \quad l=vertical,
    \end{aligned}\quad 
    C_{\theta,nl}(\theta)=\bigg\{
    \begin{aligned}
        & \sin n\theta, \quad l=horizontal,\\
        & \cos n\theta, \quad l=vertical,
    \end{aligned}
\end{equation}
being $u_{nm}$ the $m$-th solution of the equation $J_n'(u_{nm})=0$. Regarding the fourth sub-index $l$, we can distinguish between modes having a zero azimuth component of the electric field, $E_{\theta}(r,\theta=0,z)=0$ (horizontal modes) from those having a zero radial component, $E_{r}(r,\theta=0,z)=0$ (vertical modes). For further simplification, we will retain only $n=1$. Thus, we can compactly write the basis inside the holes in Dirac notation as:
\begin{equation}
\label{eq:hole basis}
    \langle \textbf{r}|f\alpha\rangle_{3D}=\begin{pmatrix}
        f^r_{\alpha}\\
        f^{\theta}_{\alpha}\\
        f^{z}_{\alpha}
    \end{pmatrix}=\begin{pmatrix}
        \frac{a}{u_{1m}r}J_1\left(u_{1m}\frac{r}{a}\right)C_{r,1l}(\theta)A_{1m}\\
        -J_1'\left(u_{1m}\frac{r}{a}\right)C_{r,1l}(\theta)A_{1m}\\
        0
    \end{pmatrix},
\end{equation}
where $\alpha$ is a multi-index denoting simultaneously the $m$-th solution, and a vertical or horizontal type of the mode ($\alpha\equiv\{m,vertical/ horizontal\}$). According to the above expressions for the electric fields, the mode $|f\alpha\rangle_{3D}$ has momentum $\nu_\alpha\equiv \nu_{1m}$, where $m$ is embedded in the multi-index $\alpha$. In the calculations, the number of considered modes is truncated up to a maximum $\max(m)$.

Due to the chosen geometry of the structure illustrated in Figure~\ref{fig:4_regions},  all boundaries between different media lie in planes parallel to the $x$-$y$ plane. Thus, we will introduce the following notations for the in-plane projections of the vector basis introduced previously in Eqs.~(\ref{eq:sp basis}, \ref{eq:oe basis}, \ref{eq:hole basis}):

\begin{equation}
    \langle \textbf{r}|sN\rangle=\frac{1}{q_N}\begin{pmatrix}
    -q_{yN}\\
    q_{xN}
    \end{pmatrix}e^{i\textbf{k}_N\textbf{r}},\quad \langle \textbf{r}|pN\rangle=\frac{1}{q_N}\begin{pmatrix}
    q_{xN}\\
    q_{yN}
    \end{pmatrix}e^{i\textbf{k}_N\textbf{r}},\quad 
    \langle \textbf{r}|f\alpha\rangle=\begin{pmatrix}
        f^r_{\alpha}\\
        f^{\theta}_{\alpha}
    \end{pmatrix},
\end{equation}

\begin{equation}
    \langle \textbf{r}|oN\rangle=\frac{1}{\chi^{o}_Nq_{N}}\begin{pmatrix}
    -q_{yN}(1-\Delta_{1N}\Delta_{zN})\\
    q_{xN}
    \end{pmatrix}e^{i\textbf{k}_N\textbf{r}},\quad \langle \textbf{r}|eN\rangle=\frac{1}{\chi^{e}_Nq_{N}}\begin{pmatrix}
    q_{xN}\frac{\Delta_{2N}-q^2_{yN}}{\Delta^e_{xN}}\\
    q_{yN}
    \end{pmatrix}e^{i\textbf{k}_N\textbf{r}},
\end{equation}
for the modes corresponding to s and p polarization, modes inside the holes, and ordinary and extraordinary modes, respectively. Remark that all these in-plane projections are normalized so that their scalar product to itself yields 1. The scalar product of two vectors is defined as:

\begin{equation}
    \langle v_1 N|v_2 N'\rangle=\int_{\Omega}d\textbf{r}\langle v_1 N|\textbf{r}\rangle\langle\textbf{r}|v_2 N'\rangle,
\end{equation}
where $\Omega$ is the area of the unit cell. Therefore, the scalar product between our vectors is defined as:

\begin{equation}
    \langle v_1 N|v_2 N'\rangle= \begin{pmatrix}
        v_{1x},v_{1y}
    \end{pmatrix} \begin{pmatrix}
        v_{2x}\\ 
        v_{2y}
    \end{pmatrix}\int^{L_y/2}_{-L_y/2}dy\int^{L_x/2}_{-L_x/2}dxe^{-i\textbf{k}_N\textbf{r}}e^{i\textbf{k}_{N'}\textbf{r}},
\end{equation}
\begin{equation}
    \langle f\alpha|f\alpha'\rangle=\int^a_0 rdr\int^{2\pi}_0d\theta\begin{pmatrix}
        f^r_{\alpha},f^{\theta}_{\alpha}
    \end{pmatrix}\begin{pmatrix}
        f^r_{\alpha'}\\
        f^{\theta}_{\alpha'}
    \end{pmatrix},
\end{equation}
\begin{equation}
    \langle s N|f\alpha\rangle= \int^a_0 rdr\int^{2\pi}_0d\theta\begin{pmatrix}
        1,\tan(-\frac{q_{xN}}{q_{yN}})
    \end{pmatrix}\begin{pmatrix}
        f^r_{\alpha'}\\
        f^{\theta}_{\alpha'}
    \end{pmatrix}
    e^{-i\textbf{k}_{N}\textbf{r}},
\end{equation}
\begin{equation}
    \langle p N|f\alpha\rangle= \int^a_0 rdr\int^{2\pi}_0d\theta\begin{pmatrix}
        1,\tan(\frac{q_{yN}}{q_{xN}})
    \end{pmatrix}\begin{pmatrix}
        f^r_{\alpha'}\\
        f^{\theta}_{\alpha'}
    \end{pmatrix}
    e^{-i\textbf{k}_{N}\textbf{r}},
\end{equation}
where $v_1,v_2\in\{s, p, o, e\}$.

With the introduced basis vectors for the electric fields, we can express the latter in each region as follows:

\begin{equation}
\label{Efield1}
    \textbf{E}_1(z)=|\sigma_0\Bar{0} 1\downarrow\rangle_{3D} e^{ik_{1z\Bar{0}}(z+d)} +\sum_{\sigma N}\bigg(R_{\sigma N}|\sigma N1\uparrow\rangle_{3D} e^{-ik_{1zN}(z+d)}\bigg) \quad z< -d,
\end{equation}
\begin{equation}
\label{Efield2}
    \textbf{E}_2(z)=\sum_{\gamma N}\bigg(A^{II}_{\gamma N}|\gamma N\downarrow\rangle_{3D} e^{ik_{\gamma zN}z}+B^{II}_{\gamma N}|\gamma N \uparrow\rangle_{3D} e^{-ik_{\gamma zN}z}\bigg)\quad -d\leq z\leq 0,
\end{equation}
\begin{equation}
\label{Efield3}
    \textbf{E}_3(z)=\sum_{\alpha}\bigg(A_{\alpha}|f\alpha\rangle_{3D} e^{i\nu_{\alpha}z}+B_{\alpha}|f\alpha\rangle_{3D} e^{-i\nu_{\alpha}z}\bigg)\quad 0< z< d_{m},
\end{equation}
\begin{equation}
\label{Efield4}
    \textbf{E}_4(z)=\sum_{\sigma N}\bigg(T_{\sigma N}|\sigma N4\downarrow\rangle_{3D} e^{ik_{4zN}(z-d_{m})}\bigg) \quad d_{m}\leq z,
\end{equation}
where $R_{\sigma N}$, $A^{II}_{\gamma N}$, $B^{II}_{\gamma N}$, $A_{\alpha}$, $B_{\alpha}$ and $T_{\sigma N}$ stand for the field amplitudes of the modes in each region. Remark that the summation in N implies in this case a double summation in $n_1$ and $n_2$. For the incident wave, the in-plane and out-of-plane momentum are $\textbf{k}_{\Bar{0}}$ and $k_{1z\Bar{0}}$, respectively, where $\Bar{0}$ denotes $\Bar{0}\equiv\{n_1=0, n_2=0\}$, and polarization is $\sigma_0$, which can be s or p-polarization. 

To unveil the mode amplitude in each region we match the fields at the boundaries of these regions. More specifically, a continuity of the in-plane components of the electric and magnetic fields must be satisfied at each of the boundaries. These boundary conditions read:

\begin{equation}
\label{elecbound}
    \textbf{E}_{1t}(z=-d)=\textbf{E}_{2t}(z=-d),\quad \textbf{E}_{2t}(z=0)=\textbf{E}_{3t}(z=0),\quad \textbf{E}_{3t}(z=d_{m})=\textbf{E}_{4t}(z=d_{m}),
\end{equation}
\begin{equation}
\begin{aligned}
    \textbf{H}_{1t}(z=-d)=\textbf{H}_{2t}(z=-d),\quad \textbf{H}_{2t}(z=0)=\textbf{H}_{3t}(z=0),\quad \textbf{H}_{3t}(z=d_{m})=\textbf{H}_{4t}(z=d_{m}),
\end{aligned}
\end{equation}

where with ¨$t$¨ we mean the in-plane vectorial component of the fields. In order to use the vector basis previously introduced, we can relate the electric and magnetic fields through the Maxwell's equation $\textbf{H}=\textbf{q}\times\textbf{E}$, with being $q$ the normalized wavevector for each of the modes. For convenience, it is worthwhile to rewrite the continuity of the magnetic field as below:

\begin{equation}
\label{magbound}
\begin{aligned}
    &-\textbf{u}_z\times\textbf{H}_1(z=-d)=-\textbf{u}_z\times\textbf{H}_2(z=-d),\\ &-\textbf{u}_z\times\textbf{H}_2(z=0)=-\textbf{u}_z\times\textbf{H}_3(z=0),\\ 
    &-\textbf{u}_z\times\textbf{H}_3(z=d_{m})=-\textbf{u}_z\times\textbf{H}_4(z=d_{m}).
\end{aligned}
\end{equation}

Thus, in order to apply the boundary conditions for the magnetic fields, we need to calculate $-\textbf{u}_z\times\textbf{q}\times\textbf{E}_i$ for all our basis vectors. It is straightforward to prove that the vectorial products of the different basis vectors can be compactly expressed as follows:

\begin{equation}
   \left(-\textbf{u}_z\times \textbf{q}\times|\sigma N\xi\uparrow,\downarrow\rangle_{3D}\right)_t=\pm Y^{\xi}_{\sigma N}|\sigma N\rangle, \quad \left(-\textbf{u}_z\times \textbf{q}\times|\gamma N\uparrow,\downarrow\rangle_{3D}\right)_t=\pm \hat{Y}^{ANI}_{N}|\gamma N\rangle,
\end{equation}
\begin{equation}
   \left(-\textbf{u}_z\times \textbf{q}\times|f\alpha\rangle_{3D}\right)_t=\pm \nu_{\alpha}|f\alpha\rangle,
\end{equation}
where +(-) stands for the modes propagating in upward "$\uparrow$" (downward "$\downarrow$" ) direction respectively; $\sigma=\{s,p\}$, $\xi=\{1,4\}$ and $\gamma=\{o,e\}$. $Y^{\xi}_{\sigma N}$ and $\hat{Y}^{ANI}_{N}$ represent the admittances of the s, p, ordinary and extraordinary modes, respectively. The admittances $Y^{\xi}_{\sigma N}$ read as:

\begin{equation}
    Y_{sN}^{\xi}=q_{\xi zN},\quad Y_{pN}^{\xi}=\frac{\varepsilon_{\xi}}{q_{\xi zN}},
\end{equation}
and $\hat{Y}^{ANI}_{N}$ is a $2\times2$ matrix:

\begin{equation}
    \hat{Y}^{ANI}_{N}=\begin{pmatrix}
    \frac{q_{xN}q_{yN}q_{ozN}\Delta_{1N}M_{poN}}{\chi^{o}_N}+q_{ozN} & \frac{\Delta_{2N}M_{poN}}{\chi^{e}_Nq_{ezN}}\\
    \frac{q_{xN}q_{yN}q_{ozN}\Delta_{1N}M_{peN}}{\chi^{o}_N} & \frac{\Delta_{2N}M_{peN}}{\chi^{e}_Nq_{ezN}}+q_{ezN}
    \end{pmatrix}.
\end{equation}

The functions $M_{po N}$ and $M_{pe N}$ are the elements of matrix $\hat{M}^{AI}_N$ , which provides the transition from the basis $\{s,p\}$ to the $\{o,e\}$ one. $M_{\sigma\gamma N}=\langle\gamma N|\sigma N\rangle$, where $\gamma\in\{o,e\}$ and $\sigma\in\{s,p\}$, thus:

\begin{equation}
    \hat{M}^{AI}_{N}=\begin{pmatrix}
    M_{soN} & M_{poN}\\
    M_{seN} & M_{peN}    
    \end{pmatrix}=\begin{pmatrix}
        P^{IA}_{soN}-M_{seN}Z^{oe}_N & P^{IA}_{poN}-M_{peN}Z^{oe}_N\\
       \frac{Z^{oe}_NP^{IA}_{poN}-P^{IA}_{peN}}{Z^{oe}_NZ^{oe}_N-1} & \frac{Z^{oe}_NP^{IA}_{poN}-P^{IA}_{peN}}{Z^{oe}_NZ^{oe}_N-1}
    \end{pmatrix}.
\end{equation}

In their turn,  $P^{IA}_{\sigma\gamma N}=\langle\sigma N|\gamma N\rangle$ functions present the elements of  matrix $\hat{P}^{IA}_N$ , which provides the transition from the basis $\{o,e\}$ to the $\{s,p\}$ one:

\begin{equation}
    \hat{P}^{IA}_{N}=\begin{pmatrix}
    P^{IA}_{soN} & P^{IA}_{seN}\\
    P^{IA}_{poN} & P^{IA}_{peN}
    \end{pmatrix}=\begin{pmatrix}
    \frac{1}{\chi^{o}_N}\bigg[1-\frac{q_{yN}^2\Delta_{1N}\Delta_{zN}}{q^2_{N}}\bigg] & \frac{1}{\chi^{e}_N}\frac{q_{xN}q_{yN}}{q^2_{N}}\bigg[1-\frac{\Delta_{2N}-q^2_{yN}}{\Delta^e_{xN}}\bigg]\\
    \frac{1}{\chi^{o}_N}\frac{q_{xN}q_{yN}\Delta_{1N}\Delta_{zN}}{q^2_{N}} & \frac{1}{\chi^{e}_N}\bigg[1+\frac{q^2_{xN}}{q^2_{N}}\bigg(\frac{\Delta_{2N}-q^2_{yN}}{\Delta^e_{xN}}-1\bigg)\bigg]
    \end{pmatrix},
\end{equation}

and $Z^{oe}_N$ is an element of the ´´projection´´ matrix $\hat{Z}_N$ for ordinary and extraordinary basis. $Z^{\gamma\gamma^\prime N}=\langle\gamma^\prime N|\gamma N\rangle$, where $\gamma,\gamma^\prime\in\{o,e\}$, thus:

\begin{equation}
    \hat{Z}_{N}=\begin{pmatrix}
    Z^{oo}_{N} & Z^{eo}_{N}\\
    Z^{oe}_{N} & Z^{ee}_{N}
    \end{pmatrix},\quad Z^{ee}_N=Z^{oo}_N=1,\quad Z^{eo}_N=Z^{oe}_N=\frac{q_{xN}q_{yN}}{Z^{ee}_NZ^{oo}_Nq^2_{N}}\bigg[1+\frac{\Delta_{2N}-q^2_{yN}}{\Delta^e_{xN}}\bigg(\Delta_{1N}\Delta_{zN}-1\bigg)\bigg].
\end{equation}

Now, we proceed with substituting the expressions for the electric field in Eqs.~(\ref{Efield1}-\ref{Efield4}) into the boundary conditions. Starting with the boundary conditions for the parallel component of the electric field, for $z=-d$, $z=0$, and $z=d_{m}$ the expressions~(\ref{elecbound}) become, respectively:

\begin{equation}
\label{eq:condition1}
    |\sigma_0\Bar{0}\rangle+\sum_{\sigma N}R_{\sigma N}|\sigma N\rangle=\sum_{\gamma N}\bigg(A^{II}_{\gamma N}e^{-}_{\gamma N}|\gamma N\rangle+B^{II}_{\gamma N}e^{+}_{\gamma N}|\gamma N\rangle\bigg),
\end{equation}
\begin{equation}
\label{eq:condition2}
    \sum_{\gamma'N}\bigg(A^{II}_{\gamma'N}|\gamma'N\rangle+B^{II}_{\gamma'N}|\gamma'N\rangle\bigg)=\sum_\alpha (A_{\alpha}+B_{\alpha})|f\alpha\rangle,
\end{equation}
\begin{equation}
\label{eq:condition3}
    \sum_\alpha (A_{\alpha}e^+_{\alpha}+B_{\alpha}e^-_{\alpha})|f\alpha\rangle=\sum_{\sigma N}T_{\sigma N}|\sigma N\rangle,
\end{equation}
where $e^{\pm}_{\gamma N}=e^{\pm idk_{\gamma zN}}$, and $e^{\pm}_{\alpha}=e^{\pm id_{m}\nu_{\alpha}}$. Concerning the boundary conditions of the magnetic field given by Eq.~(\ref{magbound}), they can now be written more explicitly in terms of the model expansion as:

\begin{equation}
\label{eq:condition4}
    Y_{\sigma_0\Bar{0}}^1|\sigma_0\Bar{0}\rangle-\sum_{\sigma N}R_{\sigma N}Y_{\sigma N}^1|\sigma N\rangle=\sum_{\gamma\gamma' N}\bigg(A^{II}_{\gamma N}e^{-}_{\gamma N}Y^{ANI}_{\gamma'\gamma N}|\gamma'N\rangle-B^{II}_{\gamma N}e^{+}_{\gamma N}Y^{ANI}_{\gamma'\gamma N}|\gamma'N\rangle\bigg),
\end{equation}

\begin{equation}
\label{eq:condition5}
    \sum_{\gamma'\gamma''N}\bigg(A^{II}_{\gamma'N}Y^{ANI}_{\gamma''\gamma'N}|\gamma''N\rangle-B^{II}_{\gamma'N}Y^{ANI}_{\gamma''\gamma'N}|\gamma''N\rangle\bigg)=\sum_\alpha (A_{\alpha}-B_{\alpha})\nu_{\alpha}|f\alpha\rangle,
\end{equation}
\begin{equation}
\label{eq:condition6}
    \sum_\alpha (A_{\alpha}e^+_{\alpha}\nu_{\alpha}-B_{\alpha}e^-_{\alpha}\nu_{\alpha})|f\alpha\rangle=\sum_{\sigma N}T_{\sigma N}Y^4_{\sigma N}|\sigma N\rangle.
\end{equation}

Projecting the Eqs.~(\ref{eq:condition1},\ref{eq:condition3}) onto $\langle \sigma N|$ and Eq.~(\ref{eq:condition2}) onto $\langle \gamma N|$, we obtain an algebraic set of equations:

\begin{equation}
\label{eq:condition7}
    \delta_{N,\Bar{0}}\delta_{\sigma,\sigma_0}+R_{\sigma N}=\sum_{\gamma}\bigg(A^{II}_{\gamma N}e^{-}_{\gamma N}P^{IA}_{\sigma\gamma N}+B^{II}_{\gamma N}e^{+}_{\gamma N}P^{IA}_{\sigma\gamma N}\bigg),
\end{equation}
\begin{equation}
\label{eq:condition8}
    Z^{\gamma\gamma'}_N\bigg(A^{II}_{\gamma'N}+B^{II}_{\gamma'N}\bigg)=\sum_\alpha (A_{\alpha}+B_{\alpha})S^{ANI}_{\gamma\alpha N},
\end{equation}
\begin{equation}
\label{eq:condition9}
    \sum_\alpha (A_{\alpha}e^+_{\alpha}+B_{\alpha}e^-_{\alpha})S^{ISO}_{\sigma\alpha N}=T_{\sigma N},
\end{equation}
where we have introduced the functions $S^{ISO}_{\sigma\alpha N}$ and $S^{ANI}_{\gamma\alpha N}$, which are the scalar products between the $\{s,p\}$ modes and the modes inside the holes; and between the $\{o,e\}$ modes and the modes inside the holes, respectively~\cite{Amitay68}:

\begin{equation}
    S^{ISO}_{pm(horizontal)N}=\langle pN|fm(horizontal)\rangle=\frac{\lambda\sqrt{2}}{\sqrt{\pi |\textbf{L}_1\times \textbf{L}_2|}}\frac{\sin(\varphi)J_1(ak_N)}{q_{N}\sqrt{u_{1m}^2-1}}\frac{q_{xN}}{q_{yN}},
\end{equation}
\begin{equation}
    S^{ISO}_{sm(horizontal)N}=\langle sN|fm(horizontal)\rangle=-\frac{\lambda\sqrt{2}}{\sqrt{\pi |\textbf{L}_1\times \textbf{L}_2|}}\frac{ak_0\cos(\varphi)J'_1(ak_N)}{[1-(ak_N/u_{1m})^2]\sqrt{u_{1m}^2-1}}\frac{q_{yN}}{q_{xN}},
\end{equation}
\begin{equation}
    S^{ISO}_{pm(vertical)N}=\langle pN|fm(vertical)\rangle=\frac{\lambda\sqrt{2}}{\sqrt{\pi|\textbf{L}_1\times \textbf{L}_2|}}\frac{\sin(\varphi)J_1(ak_N)}{q_{N}\sqrt{u_{1m}^2-1}},
\end{equation}
\begin{equation}
    S^{ISO}_{sm(vertical)N}=\langle sN|fm(vertical)\rangle=\frac{\lambda\sqrt{2}}{\sqrt{\pi |\textbf{L}_1\times \textbf{L}_2|}}\frac{ak_0\cos(\varphi)J'_1(ak_N)}{[1-(ak_N/u_{1m})^2]\sqrt{u_{1m}^2-1}},
\end{equation}

\begin{equation}
    S^{ANI}_{\gamma\alpha N}=P^{IA}_{s\gamma N}S^{ISO}_{s\alpha N}+P^{IA}_{p\gamma N}S^{ISO}_{p\alpha N},
\end{equation}
where $\varphi=\arctan(q_{yN}/q_{xN})$. The functions $S^{ANI}_{\gamma\alpha N}$ can be compactly written in a matrix form, namely $\hat{S}^{ISO}_N$ and $\hat{S}^{ANI}_N$, defined as:

\begin{equation}
    \hat{S}^{ISO}_{N}=\begin{pmatrix}
        S^{ISO}_{s\{1(vert)\}N} & S^{ISO}_{s\{1(horiz)\}N} & S^{ISO}_{s\{2(vert)\}N} & \cdots & S^{ISO}_{s\{max(m)(vert)\}N} & S^{ISO}_{s\{max(m)(horiz)\}N}\\
        S^{ISO}_{p\{1(vert)\}N} & S^{ISO}_{p\{1(horiz)\}N} & S^{ISO}_{p\{2(vert)\}N} & \cdots & S^{ISO}_{p\{max(m)(vert)\}N} & S^{ISO}_{p\{max(m)(horiz)\}N}
    \end{pmatrix},
\end{equation}
where $vert\equiv vertical$ and $horiz\equiv horizontal$, and $\hat{S}^{ANI}_N=\left(\hat{P}^{IA}_{N}\right)^\dagger\hat{S}^{ISO}_{N}$.

Concerning Eqs.~(\ref{eq:condition4}-\ref{eq:condition6}), we ''project'' each equation using a basis that fulfills the continuity of the parallel component of the magnetic field, which is continuous everywhere on the interface $z=-d$. Thus, we will project Eq.~(\ref{eq:condition4})  using, not the waveguide modes, but the modes in bulk $\langle \sigma N|$ (as the magnetic field is continuous everywhere on the interface $z=-d$), while the Eqs.~(\ref{eq:condition5},\ref{eq:condition6}), we will multiply by $\langle f\alpha|$, as at the interface $z=0$ the magnetic field is continuous only in the areas delimited by the holes. By following this procedure, we obtain the following three equations:

\begin{equation}
\label{eq:condition10}
    Y_{\sigma_0\Bar{0}}^1\delta_{N,\Bar{0}}\delta_{\sigma,\sigma_0}-Y^1_{\sigma N}R_{\sigma N}=\sum_{\gamma\gamma'}\bigg(A^{II}_{\gamma N}e^{-}_{\gamma N}Y^{ANI}_{\gamma'\gamma N}P^{IA}_{\sigma\gamma'N}-B^{II}_{\gamma N}e^{+}_{\gamma N}Y^{ANI}_{\gamma'\gamma N}P^{IA}_{\sigma\gamma'N}\bigg),
\end{equation}

\begin{equation}
\label{eq:condition11}
    \sum_{\gamma'\gamma''N}\bigg(A^{II}_{\gamma'N}Y^{ANI}_{\gamma''\gamma'N}-B^{II}_{\gamma'N}Y^{ANI}_{\gamma''\gamma'N}\bigg)S^{ANI}_{\gamma''\alpha N}=(A_{\alpha}-B_{\alpha})\nu_{\alpha},
\end{equation}

\begin{equation}
\label{eq:condition12}
    A_{\alpha}e^+_{\alpha}\nu_{\alpha}-B_{\alpha}e^-_{\alpha}\nu_{\alpha}=\sum_{\sigma N}T_{\sigma N}Y^4_{\sigma N}S^{ISO}_{\sigma\alpha N}.
\end{equation}

We have thus derived the linear system of equations, presented in Eqs.~(\ref{eq:condition7}, \ref{eq:condition8}, \ref{eq:condition9}, \ref{eq:condition10}, \ref{eq:condition11}, \ref{eq:condition12}) for the unknown amplitudes of the fields $R_{\sigma N}$, $A^{II}_{\gamma N}$, $B^{II}_{\gamma N}$, $A_{\alpha}$, $B_{\alpha}$ and $T_{\sigma N}$. For compactness, we will translate the system into matrix format. For this purpose, we introduce the following matrix notation:

\begin{gather}
    \hat{e}^{\pm}=\begin{pmatrix}
        e^{\pm}_{\{1,vertical\}} & 0 & 0 &\cdots & 0 & 0\\
        0 & e^{\pm}_{\{1,horizontal\}} & 0 & \cdots & 0 & 0\\
        0 & 0 & e^{\pm}_{\{2,vertical\}} & \cdots & 0 & 0\\
         & & & & & \\
        \vdots & \vdots & & \ddots & \vdots & \vdots\\
        & & & & & \\
        0 & 0 & 0 & \cdots & e^{\pm}_{\{max(m),vertical\}} & 0\\
        0 & 0 & 0 & \cdots & 0 & e^{\pm}_{\{max(m),horizontal\}}
    \end{pmatrix},\\
    \hat{e}^{\pm}_N=\begin{pmatrix}
        e^{\pm}_{oN} & 0\\
        0 & e^{\pm}_{eN}
    \end{pmatrix}, \quad\hat{Y}^\xi_G=\begin{pmatrix}
        Y^\xi_{Gs} & 0 \\
        0 & Y^\xi_{Gp}
    \end{pmatrix},\\
     \hat{Y}_h=\begin{pmatrix}
        \nu_{\{1,vertical\}} & 0 & 0 &\cdots & 0 & 0\\
        0 & \nu_{\{1,horizontal\}} & 0 & \cdots & 0 & 0\\
        0 & 0 & \nu_{\{2,vertical\}} & \cdots & 0 & 0\\
         & & & & & \\
        \vdots & \vdots & & \ddots & \vdots & \vdots\\
        & & & & & \\
        0 & 0 & 0 & \cdots & \nu_{\{max(m),vertical\}} & 0\\
        0 & 0 & 0 & \cdots & 0 & \nu_{\{max(m),horizontal\}}
    \end{pmatrix},
\end{gather}

where $\hat{e}^{\pm}_N$ and $\hat{e}^{\pm}$ are diagonal matrices containing the exponential decay through the anisotropic slab and the hole, respectively; $\hat{Y}_h$ is the admittance matrix of the modes inside the holes, and $\hat{Y}^{\xi}_N$ is a diagonal matrix with the admittance of the s and p modes. These matrices allow us to rewrite the system of equations in Eqs.~(\ref{eq:condition7},\ref{eq:condition8},\ref{eq:condition9},\ref{eq:condition10},\ref{eq:condition11},\ref{eq:condition12}) in the following way:

\begin{equation}
\label{eq:condition13}
    \textbf{R}_N=-\begin{pmatrix}
        \delta_{N,\Bar{0}}\delta_{s,\sigma_0}\\
        \delta_{N,\Bar{0}}\delta_{p,\sigma_0}
    \end{pmatrix}+\hat{P}^{IA}_{N}\hat{e}^-_{N}\textbf{A}^{II}_N+\hat{P}^{IA}_{N}\hat{e}^+_{N}\textbf{B}^{II}_N,
\end{equation}
\begin{equation}
\label{eq:condition14}
    \textbf{Y}^{1}_{\Bar{0}}-\hat{Y}^{1}_{N}\textbf{R}_N=\hat{P}^{IA}_{N}\hat{Y}^{ANI}_N\hat{e}^-_{N}\textbf{A}^{II}_N-\hat{P}^{IA}_{N}\hat{Y}^{ANI}_N\hat{e}^+_{N}\textbf{B}^{II}_N,
\end{equation}
\begin{equation}
\label{eq:condition15}
    \hat{Z}_N(\textbf{A}^{II}_N+\textbf{B}^{II}_N)=\hat{S}^{ANI}_N(\textbf{A}+\textbf{B}),
\end{equation}
\begin{equation}
\label{eq:condition16}
    \sum_N\bigg((\hat{S}^{ANI}_N)'\hat{Y}^{ANI}_N\textbf{A}^{II}_N-(\hat{S}^{ANI}_N)'\hat{Y}^{ANI}_N\textbf{B}^{II}_N\bigg)=\hat{Y}_h(\textbf{A}-\textbf{B}),
\end{equation}
\begin{equation}
\label{eq:condition17}
    \hat{S}^{ISO}_N(\hat{e}^+\textbf{A}+\hat{e}^-\textbf{B})=\textbf{T}_{N},
\end{equation}
\begin{equation}
\label{eq:condition18}
    \hat{Y}_h (\hat{e}^+\textbf{A}-\hat{e}^-\textbf{B})=\sum_{N}(\hat{S}^{ISO}_{N})'\hat{Y}^4_{N}\textbf{T}_{N},
\end{equation}
where $\textbf{R}_N$, $\textbf{A}^{II}_N$, $\textbf{B}^{II}_N$, and $\textbf{T}_{N}$ are two-components vectors:

\begin{equation}
    \textbf{R}_N=\begin{pmatrix}
        R_{sN}\\
        R_{pN}
    \end{pmatrix},\quad \textbf{A}^{II}_N=\begin{pmatrix}
        A^{II}_{oN}\\
        A^{II}_{eN}
    \end{pmatrix},\quad \textbf{B}^{II}_N=\begin{pmatrix}
        B^{II}_{oN}\\
        B^{II}_{eN}
    \end{pmatrix},\quad \textbf{T}_N=\begin{pmatrix}
        T_{sN}\\
        T_{pN}
    \end{pmatrix},
\end{equation}
and $\textbf{A}$, $\textbf{B}$ are $max(m)$-components vectors, ordered as:

\begin{equation}
    \textbf{A}=\begin{pmatrix}
        A_{\{1,vertical\}}\\
        A_{\{1,horizontal\}}\\
        A_{\{2,vertical\}}\\
        \vdots\\
        A_{\{max(m),vertical\}}\\
        A_{\{max(m),horizontal\}}
    \end{pmatrix},\quad \textbf{B}=\begin{pmatrix}
        B_{\{1,vertical\}}\\
        B_{\{1,horizontal\}}\\
        B_{\{2,vertical\}}\\
        \vdots\\
        B_{\{max(m),vertical\}}\\
        B_{\{max(m),horizontal\}}
    \end{pmatrix}.
\end{equation}

Also, $\textbf{Y}^{1}_{\Bar{0}}$ is a two-components vector defined by the following product:

\begin{equation}
    \textbf{Y}^{1}_{\Bar{0}}=\begin{pmatrix}
        Y^1_{sN} & 0 \\
        0 & Y^1_{pN}
    \end{pmatrix}\begin{pmatrix}
        \delta_{N,\Bar{0}}\delta_{s,\sigma_0}\\
        \delta_{N,\Bar{0}}\delta_{p,\sigma_0}
    \end{pmatrix}.
\end{equation}

Substituting the Eq.~(\ref{eq:condition13}) into Eq.~(\ref{eq:condition14}), we obtain:

\begin{equation}
    2\textbf{Y}^{1}_{\Bar{0}}=(\hat{M}^H_{NrA}+\hat{Y}^{1}_{N}\hat{M}^E_{NrA})\textbf{A}^{II}_N+(-\hat{M}^H_{NrB}+\hat{Y}^{1}_{N}\hat{M}^E_{NrB})\textbf{B}^{II}_N,
\end{equation}
where $\hat{M}^E_{NrA}=\hat{P}^{IA}_{N}\hat{e}^-_{N}$, $\hat{M}^E_{NrB}=\hat{P}^{IA}_{N}\hat{e}^+_{N}$, $\hat{M}^H_{NrA}=\hat{P}^{IA}_{N}\hat{Y}^{ANI}_N\hat{e}^-_{N}$, and $\hat{M}^H_{NrB}=\hat{P}^{IA}_{N}\hat{Y}^{ANI}_N\hat{e}^+_{N}$. From there, we can obtain the expression for $\textbf{B}^{II}_N$:

\begin{equation}
\label{eq:condition19}
   \textbf{B}^{II}_N=\hat{M}_{NB}^{-1}(2\textbf{Y}^{1}_{\Bar{0}}-\hat{M}_{NA}\textbf{A}^{II}_N), 
\end{equation}
with $\hat{M}_{NA}=\hat{M}^H_{NrA}+\hat{Y}^{1}_{N}\hat{M}^E_{NrA}$ and $\hat{M}_{NB}=-\hat{M}^H_{NrB}+\hat{Y}^{1}_{N}\hat{M}^E_{NrB}$. Substituting the Eq.~(\ref{eq:condition19}) into Eq.~(\ref{eq:condition15}), the following expression for the field amplitude $\textbf{A}^{II}_N$ can be obtained:

\begin{equation}
\label{eq:condition20}
    \textbf{A}^{II}_N=-\Tilde{\textbf{I}}_N+\hat{\Xi}_N\hat{S}^{ANI}_N(\textbf{A}+\textbf{B}),
\end{equation}
where $\hat{\Xi}_N=[\hat{Z}_G(\hat{I}_2-\hat{M}_{NB}^{-1}\hat{M}_{NA})]^{-1}$, $\Tilde{\textbf{I}}_N=2\hat{\Xi}_N\hat{Z}_N\hat{M}_{NB}^{-1}\textbf{Y}^{1}_{\Bar{0}}$, and $\hat{I}_2$ is the $2\times 2$ identity matrix. Now, if we substitute the expressions for the field amplitudes $\textbf{A}^{II}_N$ and $\textbf{B}^{II}_N$ from Eqs.~(\ref{eq:condition19},\ref{eq:condition20}) into Eq.~(\ref{eq:condition16}) we have:

\begin{equation}
    \textbf{I}^{REN}+\hat{G}_{up}(\textbf{A}+\textbf{B})=\hat{Y}_h(\textbf{A}-\textbf{B}),
\end{equation}
where $\textbf{I}^{REN}$ and $\hat{G}_{up}$ are defined as below:

\begin{equation}
    \textbf{I}^{REN}=\sum_N\bigg((\hat{S}^{ANI}_N)'\hat{Y}^{ANI}_N(-2\hat{M}_{NB}^{-1}\textbf{Y}^{1}_{\Bar{0}}-(\hat{I}_2+\hat{M}_{NB}^{-1}\hat{M}_{NA})\Tilde{\textbf{I}}_N)\bigg),
\end{equation}
\begin{equation}
    \hat{G}_{up}=\sum_N\bigg((\hat{S}^{ANI}_N)'\hat{Y}^{ANI}_N(\hat{I}_2+\hat{M}_{NB}^{-1}\hat{M}_{NA})\hat{\Xi}_N\hat{S}^{ANI}_N\bigg).
\end{equation}

Finally, if we substitute Eq.~(\ref{eq:condition17}) into Eq.~(\ref{eq:condition18}) we obtain:

\begin{equation}
    \textbf{0}=(\hat{Y}_h-\hat{G}_{low})\hat{e}^+\textbf{A}+(-\hat{Y}_h-\hat{G}_{low})\hat{e}^-\textbf{B},
\end{equation}
where $\textbf{0}=(0,\cdots,0)^T$ is $max(m)$-component vector, and $\hat{G}_{low}=\sum_{N}(\hat{S}^{ISO}_{N})'\hat{Y}^4_{N}\hat{S}^{ISO}_N$. Thus, the final linear system of equations for the amplitudes of the fields of the modes inside the holes reads as:

\begin{equation}
\label{eq:pre-final_expression}
\begin{aligned}
    &\textbf{I}^{REN}=(\hat{Y}_h-\hat{G}_{up})\textbf{A}+(-\hat{Y}_h-\hat{G}_{up})\textbf{B},\\
    &\textbf{0}=(\hat{Y}_h-\hat{G}_{low})\hat{e}^+\textbf{A}+(-\hat{Y}_h-\hat{G}_{low})\hat{e}^-\textbf{B}.
\end{aligned}
\end{equation}

After some straightforward algebraic calculations, the resulting system of equations closely resembles the one obtained in previous studies of extraordinary optical transmission (in simpler geometries). As we see, in the system of equations only the fields inside the holes appear~\cite{Martin2008}.

Equations~(\ref{eq:pre-final_expression}) can also be compactly written in the matrix form:

\begin{equation}
    \label{eq:final_expression}
    \boxed{\textbf{I}^{TOT}=\hat{D}_{TOT}\textbf{AB}^{TOT},}
\end{equation}
\begin{equation}
    \textbf{I}^{TOT}=\begin{pmatrix}
        \textbf{I}^{REN}\\
        \textbf{0}
    \end{pmatrix}, \quad \textbf{AB}^{TOT}=\begin{pmatrix}
        \textbf{A}\\
        \textbf{B}
    \end{pmatrix}, \quad \hat{D}_{TOT}=\begin{pmatrix}
        \hat{Y}_h-\hat{G}_{up} & -\hat{Y}_h-\hat{G}_{up}\\
        (\hat{Y}_h-\hat{G}_{low})\hat{e}^+ & (-\hat{Y}_h-\hat{G}_{low})\hat{e}^-
    \end{pmatrix}.
\end{equation}
Once the field vectors $\textbf{A}$ and $\textbf{B}$ are calculated, the spatial field distributions in each region can be reconstructed by substituting them into Eqs.~(\ref{eq:condition17}, \ref{eq:condition13}, \ref{eq:condition19}, \ref{eq:condition20}).

Importantly, our method allows one to determine the contribution of each mode in the total field distribution. In addition, the dispersion relation, characterizing the behavior of waves within the structure, is determined by the roots of the determinant: $|\hat{D}_{TOT}|=0$.

The linear system described by Equation~(\ref{eq:final_expression}) exhibits an infinite number of field harmonics $N$ and hole modes $\alpha$. For numerical solution, truncation is necessary up to a specified order $N_{max}$ and $\max(m)$, where $N_{max}$ means that $n_1, n_2\in\{-N_{max},\ldots,0,\ldots, N_{max}\}$. Both $N_{max}$ and $\max(m)$ have to be sufficiently large to ensure convergence of the solution, although yet small enough to maintain a reasonable computation time. In this article, the employed truncation values of $N_{max}=12$ and $\max(m)=5$, retaining both horizontal and vertical modes for the holes.

\section{Representing the far-field response and band structure of the twisted PC}

The numerical solution of the system of equations yields the field vectors $\textbf{A}$ and $\textbf{B}$. Substituting these vectors into Eqs.~(\ref{eq:condition13}, \ref{eq:condition19}, \ref{eq:condition20}), we obtain the field Fourier harmonics amplitudes $R_{\sigma N}$. The spectra for normal incidence depicted in Figure 2~(b, f, j) are obtained by setting $k_x=k_y=0$ in the incident light, with the reflection coefficient given by the following formula~\cite{Nikitin17}:

\begin{equation}
    R=\sum_{\sigma N} Re\bigg(\frac{Y^1_{\sigma N}}{Y^1_{\sigma \Bar{0}}}\bigg)|R_{\sigma N}|^2.
\end{equation}
The real field patterns shown in Figure 2~(c, g, k) are reconstructed by substituting the obtained field Fourier harmonics amplitudes, $R_{\sigma N}$, into Eq.~(\ref{Efield1}). The false colorplots resembling the isofrequency curves displayed in Figure 2~(d, h, l) and Figure 3~(i, j, k, l) are generated by sweeping in $k_x$ and $k_y$ at a fixed frequency and plotting $\sum_{\sigma N}|R_{\sigma N}|^2$. A similar procedure is employed to generate the colorplots which depict features of the bandstructure in Figure 3~(d, e, f, g, h), where the sweeping in $k_x$ and $k_y$ follows the traces indicated in Figure 3c by dashed blue lines, for each frequency.

Regarding the colored dashed lines in Figure 2~(d, h, l) and Figure 3~(i, j, k, l), as well as the grey lines in Figure 3d, they represent the roots ($|F(k_xN,k_yN)|=0$) of the implicit equation:

\begin{equation}
\label{eq:empty_grating}
    F(k_{xN},k_{yN})=k_N-F_{m}(k_{xN},k_{yN}),
\end{equation}
where $F_{m}$ denotes the momentum of polaritons in a slab of $\alpha\textrm{-MoO}_3$ atop a bare gold layer. This expression can be derived from Eq.~(71) of the article~\cite{AlvarezPerez19}, assuming the thickness of the anisotropic slab to be double (due to symmetry, as gold acts as a mirror), with the superstate equal to the substrate (also due to symmetry), and setting $l=1$. The final expression for $F_{m}(k_{xN},k_{yN})$ reads as follows:

\begin{equation}
    F_{m}(k_{xN},k_{yN})=\frac{\rho(k_{xN},k_{yN})}{2d}\bigg[2\arctan\bigg(\frac{\varepsilon_1\rho}{\varepsilon_z}\bigg)+\pi\bigg],
\end{equation}
where $\rho(k_{xN},k_{yN})$ is defined as:

\begin{equation}
    \rho(k_{xN},k_{yN})=i\sqrt{\frac{\varepsilon_zk_N^2}{\varepsilon_xk_{xN}^2+\varepsilon_yk_{yN}^2}}.
\end{equation}

Since it is not possible to obtain an explicit expression from Equation~(\ref{eq:empty_grating}), we opted to generate a colorplot for each combination of $n_1$ and $n_2$. We then isolate the curves generated by the minima of the function $|F(k_xN,k_yN)|$ and overlay them on the plots, employing distinct coloration or line styles to differentiate between modes $(n_1,n_2)$.

\section{Extracting the momentum of polaritons from the experimental data}
 
\begin{figure}[h]
    \centering
    \includegraphics[scale=0.5]{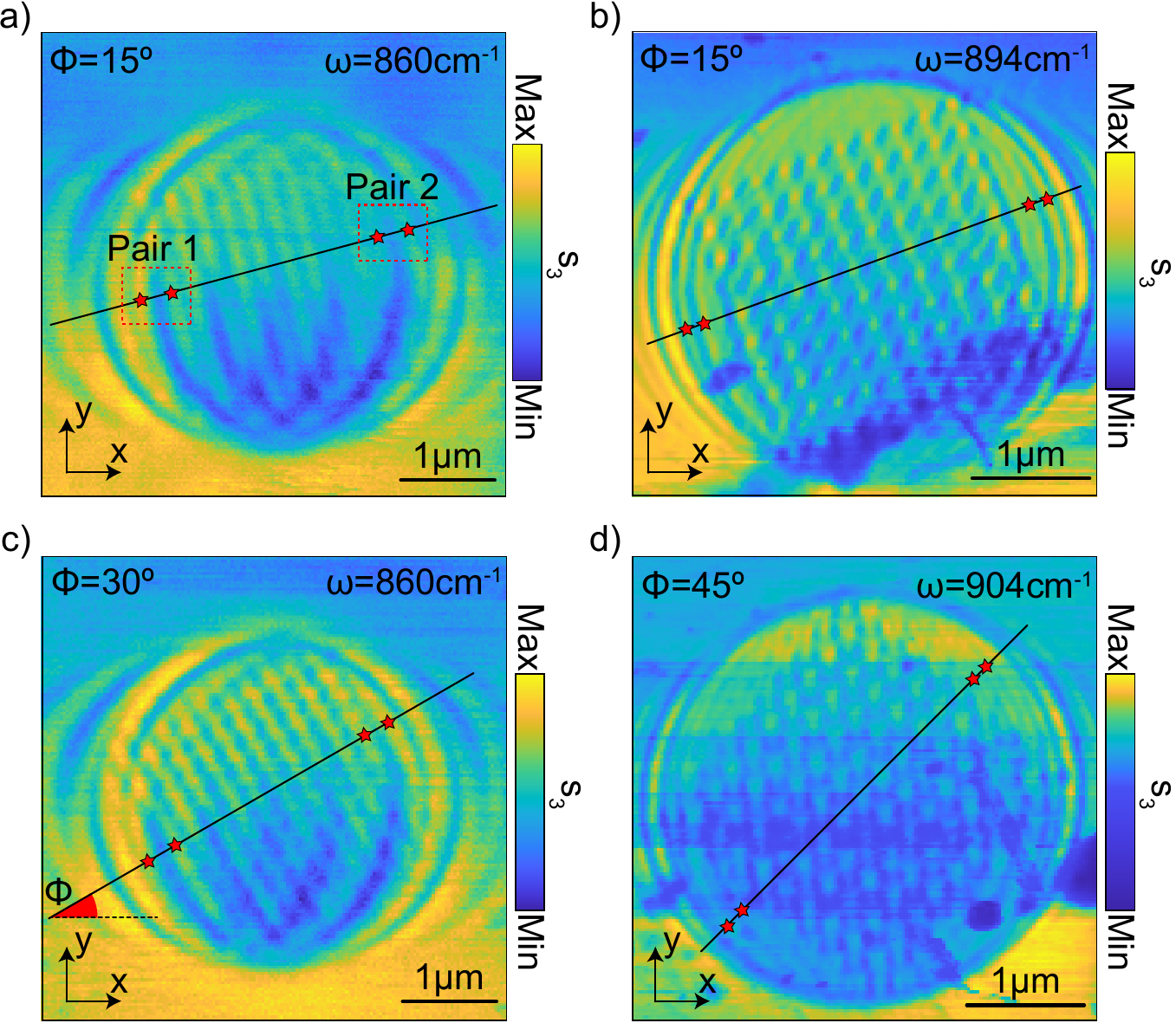}
    \caption{(a-d) Colorplot of the third signal harmonic, $\textrm{s}_3$ , from the s-SNOM experiment for twist angles of $15\grad$, $15\grad$, $30\grad$ and $45\grad$ at the frequency of $860\cm^{-1}$, $882.2\cm^{-1}$, $875.7\cm^{-1}$, and $864.1\cm^{-1}$, respectively. The black line is oriented parallel to the $\textbf{L}_1$ lattice vector, thus along the $\Gamma\rightarrow \textrm{X}$ direction in the reciprocal space. The red stars indicate the position of the maxima taken to calculate the PhP wavelength (momentum).}
    \label{fig:momentum_extraction}
\end{figure}

\begin{figure}[H]
    \centering
    \includegraphics[scale=0.4]{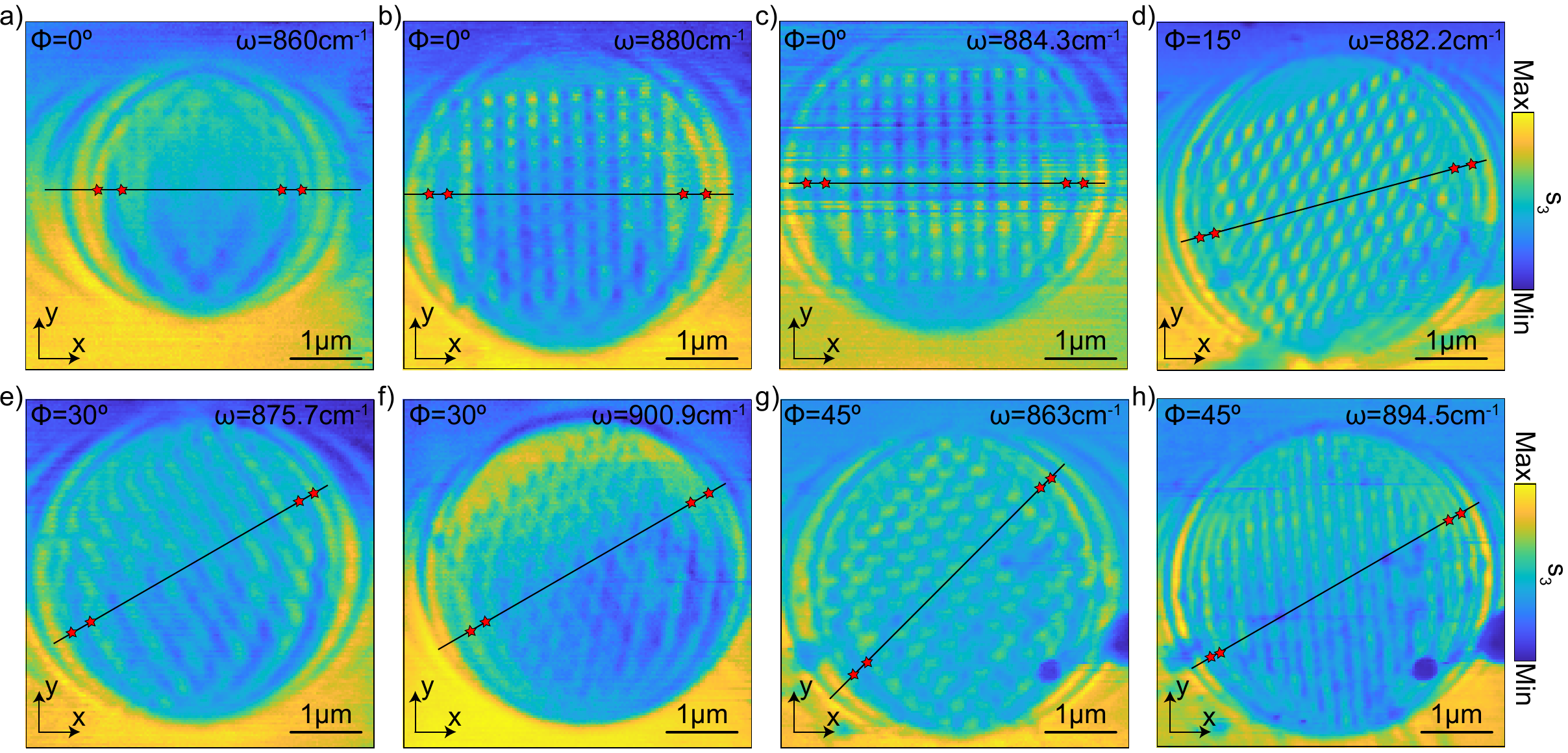}
    \caption{(a-h) Colorplot of the third signal harmonic, $\textrm{s}_3$, from the s-SNOM experiment to extract the experimental points for the bandstructure in the main text in Figure 3. The twist angle is indicated in the upper left corner and the frequency in the upper right corner. The black line is oriented parallel to the $\textbf{L}_1$ lattice vector, thus along the $\Gamma\rightarrow \textrm{X}$ direction in the reciprocal space. The red stars indicate the position of the maxima taken to calculate the PhP wavelength (momentum).}
    \label{fig:rest_momentum_extraction}
\end{figure}

In this section, we will explain how to calculate the experimental points used to reconstruct the band structure in Figure 3e-h of the main text. In Figure~\ref{fig:momentum_extraction} we present the near-field patterns for different configurations of angles and frequencies, which were used to obtain the momentum in Figure 3e-h of the main text. To extract the PhP momentum, the HAs were surrounded by a drilled ring. This design introduces a boundary in all in-plane directions, which launches PhPs. The distance between the fringes of these PhPs is measured to determine their momentum. To do so, we draw a line oriented along the $\textbf{L}_1$ lattice vector, thus forming an angle $\Phi$ with the x-axis, as depicted in Figure~\ref{fig:momentum_extraction}c. Afterward, we select the crossing point of the black line with the maxima of the field pattern from the launched polariton by the edge of the ring within the interior of the hole array, as indicated by the red stars. Due to the symmetry of the HA, the line intersects two edges of the ring simultaneously, allowing us to group fringes based on their proximity to the edges, as for instance in Figure~\ref{fig:momentum_extraction}a, where there are two groups of points, pair 1 and pair 2. Subsequently, we measure the distance between adjacent selected points within the same group and calculate the average of all the distances, denoting this value as $\lambda_{exp}$. This $\lambda_{exp}$ is the wavelength of the polaritons along the $\Gamma\rightarrow \textrm{X}$ direction since they are extracted along a line parallel of the $\textbf{L}_1$ vector. By calculating $G_{exp}=2\pi/\lambda_{exp}$ we obtain the experimental momentum $G_{exp}$ of the polariton along that direction. Generally, $G_{exp}$ is larger than the modulus of the vector $\textbf{g}_1/2$, thus falling inside the second BZ. We translate the extracted momentum from the second BZ to the first BZ by subtracting $g_1-G_{exp}$, and plot the resulting momentum in the band structure by blue asterisk symbols (Figure 3e-h). All the experimental measurements used to extract the experimental momentum can be found in Figure~\ref{fig:momentum_extraction} and Figure~\ref{fig:rest_momentum_extraction}.

Without loss of generality, we can repeat this procedure to obtain the momentum along any other in-plane direction, thus reciprocal space direction. 

\section{Near-field patterns for different Bragg resonance frequencies}

\begin{figure}[H]
    \centering
    \includegraphics[scale=0.4]{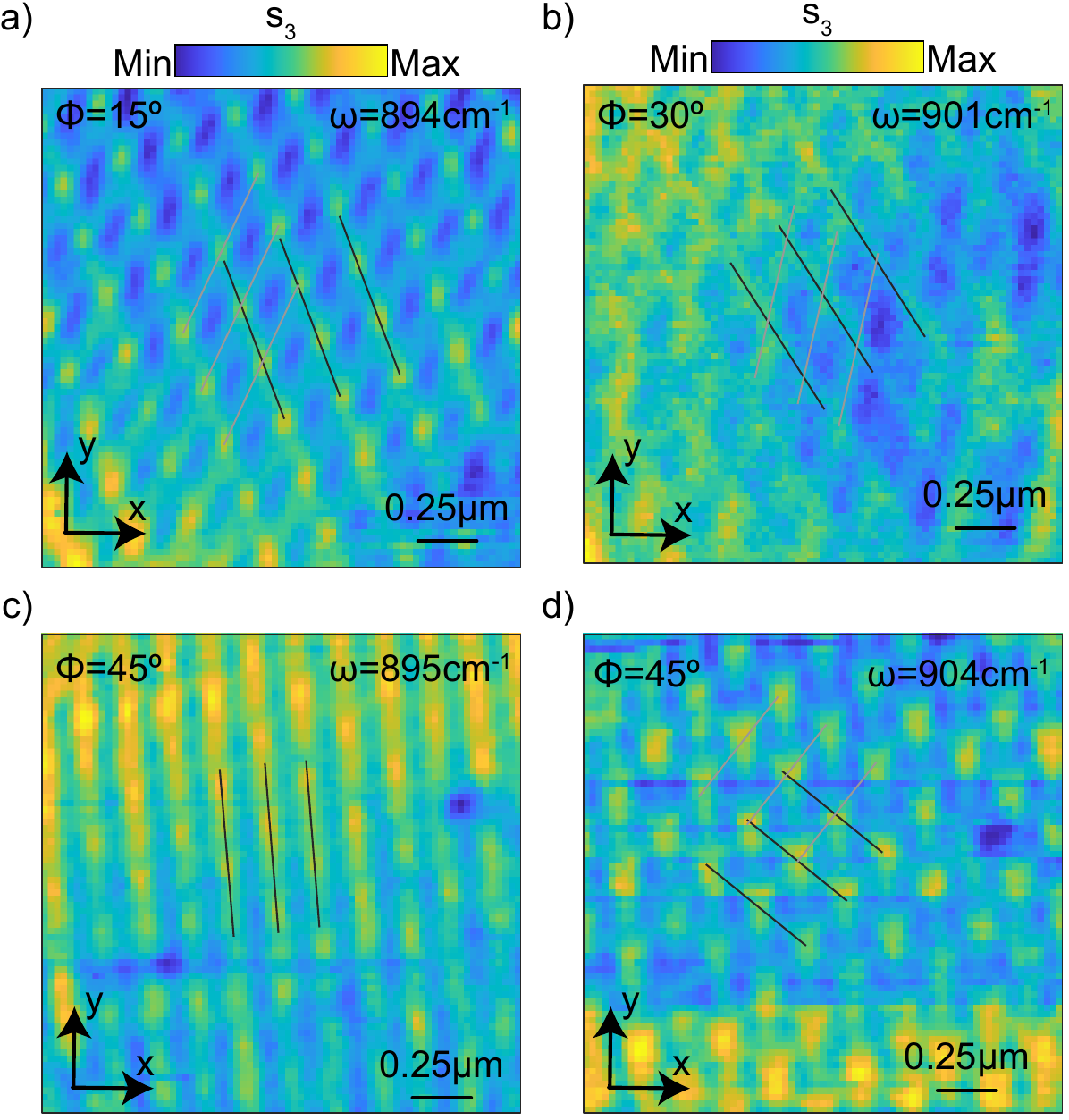}
    \caption{(a-d) Colorplot of the $\textrm{s}_3$ harmonic from the s-SNOM scanned experiment for twist angles of $15\grad$, $30\grad$, $45\grad$ and $45\grad$ at a frequency of $894\cm^{-1}$, $901\cm^{-1}$, $895\cm^{-1}$, and $904\cm^{-1}$, respectively. Solid black and gray lines indicating the position of the fringes are plotted as a guide to the eye.}
    \label{fig:diff_fres_angles}
\end{figure}

In this section, we show additional s-SNOM data showing fringes associated with Bloch modes. In Figure~\ref{fig:diff_fres_angles} we illustrate different configurations of angles and frequencies within Bragg resonance condition for different orders out of $(\pm1, 0)$ order mode. Remark that the Bragg resonance order for each case is different. As a guide to the eye, solid black and gray lines are drawn indicating the position and orientation of the fringes.

\newpage



\begin{thebibliography}{16} \addcontentsline{toc}{section}{References}

\bibitem{AlvarezPerez19}
Gonzalo \'Alvarez-P\'erez et al. "Analytical approximations for the dispersion of electromagnetic modes in slabs of biaxial crystals". In: \textit{Phys. Rev. B} 100, (23 2019), p. 235408. DOI: 10.1103/PhysRevB.100.235408.

\bibitem{Amitay68}
N. Amitay and V. Galindo. "On the scalar product of certain circular and cartesian wave functions". In: \textit{IEEE Trans. Microwave Theory Tech.} 16 (4 1968), pp. 265-266.

\bibitem{Martin2010}
F. J. Farcia-Vidal et al. "Light passing through subwavelength apertures". In: \textit{Rev. Mod. Phys.} 82 (1 2010), pp. 729-787. DOI: 10.1103/RevModPhys.82.729.

\bibitem{Martin2008}
L. Mart\'in-Moreno and F. J. Garc\'ia-Vidal. "Minimal model for optical transmission through holey metal films". In: \textit{Journal of Physics: Condensed Matter} 20.30 (2008), p. 304214. DOI: 10.1088/0953-8984/20/30/304214.

\bibitem{Nikitin17}
Alexey Yu Nikitin. \textit{Graphene Plasmonics. World Scientific Handbook of Metamaterials and Plasmonics. World Scientific Series in Nanoscience and Nanotechnology.} Vol. 4. Cambridge University Press, 2017. Chap. 8, pp.307-338.

\bibitem{Roberts87}
A. Roberts. "Electromagnetic theory of diffraction by a circular aperture in a thick, perfectly conducting screen". In: \textit{J. Opt. Soc. Am.} 4 (1987), pp. 1970-1983.

\end{thebibliography}
\end{document}